\documentclass[pdflatex,sn-mathphys-num]{sn-jnl}

\usepackage{graphicx}
\usepackage{multirow}
\usepackage{amsmath,amssymb,amsfonts}
\usepackage{amsthm}
\usepackage{mathrsfs}
\usepackage[title]{appendix}
\usepackage{xcolor}
\usepackage{textcomp}
\usepackage{booktabs}
\usepackage{array}
\usepackage{enumitem}

\usepackage{tabularx}
\usepackage{longtable}

\newcolumntype{Y}{>{\raggedright\arraybackslash}X}

\providecommand{\biwmon}[1]{\ensuremath{m_{\scriptscriptstyle #1}}}
\providecommand{\biwthree}[3]{%
\ensuremath{\biwmon{#1}\allowbreak+\biwmon{#2}\allowbreak+\biwmon{#3}}}

\theoremstyle{thmstyleone}
\newtheorem{theorem}{Theorem}[section]
\newtheorem{proposition}[theorem]{Proposition}
\newtheorem{lemma}[theorem]{Lemma}
\newtheorem{corollary}[theorem]{Corollary}

\theoremstyle{thmstyletwo}
\newtheorem{remark}[theorem]{Remark}

\theoremstyle{thmstylethree}

\newtheorem{construction}[theorem]{Construction}

\DeclareMathOperator{\wt}{wt}
\DeclareMathOperator{\supp}{supp}
\newcommand{\RM}{\mathrm{RM}}

\begin{document}

\title[Weight spectrum of binary Reed--Muller codes]{Some Results on the Weight Spectrum of Binary Reed--Muller Codes \(\RM(m-7,m)\)}

% Replace the placeholder author and affiliation information before submission.
% \author*[1]{\fnm{Authors} \sur{}}
\author[1]{\fnm{Zibo} \sur{Wei}}

\author[1]{\fnm{Bo} \sur{Zhou}}

\author[3,4]{\fnm{Xiao} \sur{Zhang}}
% \email{xiao.zh@buaa.edu.cn}

\author*[2]{\fnm{Hao} \sur{Chen}}
\email{haochen@jnu.edu.cn}

\affil[1]{\orgdiv{School of Artificial Intelligence},
\orgname{Beihang University},
\orgaddress{\city{Beijing},
\postcode{100191},
\country{China}}}

\affil*[2]{\orgdiv{College of Information Science and Technology},
\orgname{Jinan University},
\orgaddress{\city{Guangzhou},
\postcode{510632},
\state{Guangdong},
\country{China}}}

\affil[3]{\orgdiv{School of Mathematical Sciences},
\orgname{Beihang University},
\orgaddress{\city{Beijing},
\postcode{100191},
\country{China}}}

\affil[4]{\orgdiv{Key Laboratory of Mathematics, Informatics and Behavioral
Semantics (LMIB)},
\orgname{Beihang University},
\orgaddress{\city{Beijing},
\postcode{100191},
\country{China}}}

\abstract{\unboldmath
We study the set of weights occurring in the high-order binary Reed--Muller
codes \(\RM(m-7,m)\).  For every \(m\ge14\), we prove that the weight spectrum
contains all even weights in \([336,2^m-336]\), together with an explicit
finite set of low weights and their complements.  The proof first establishes
the corresponding base-code inclusion in \(\RM(7,14)\).  This base step uses
four-block concatenation, explicit degree-six monomial witnesses, and known
spectrum inputs for \(\RM(5,10)\) and \(\RM(6,12)\).  The base inclusion is then
lifted to all \(m\ge14\) by a concatenation-based induction.  We also analyze 
the remaining base-code target weights \(322,326,330,334\) in several restricted 
families, including four-block branches, affine-flat and structured inner 
corrections in a fixed Carlet-type Maiorana--McFarland outer shell, and sparse sums 
of degree-\(7\) coordinate monomials. The residual analysis delineates
several construction routes that cannot yield the four unresolved
base-code weights, thereby sharpening the remaining \(c=7\) spectrum
problem.
}



\maketitle

\section{Introduction}
\label{sec:introduction}

Let \(0\le r\le m\).  The binary Reed--Muller code \(\RM(r,m)\) is the set of
truth tables of all \(m\)-variable Boolean functions of algebraic degree at most
\(r\).  We write
\[
        W(r,m)=\{\wt(f)\mid f\in\RM(r,m)\}
\]
for its weight spectrum, namely the set of distinct Hamming weights occurring
in the code.  In this paper we study the spectrum as a set, not the full weight
distribution.

The low-weight spectrum of Reed--Muller codes has been studied classically.
Kasami and Tokura determined the first low-weight range, and Kasami, Tokura
and Azumi extended the classification to all weights below \(5d/2\), where
\(d=2^{m-r}\) is the minimum distance of \(\RM(r,m)\)
\cite{Kasami1970Weight,Kasami1976WeightEnumeration}.  Complete weight spectra
are substantially harder and are known in explicit form only for a limited
number of special orders, including \(r=0,1,2,m-2,m-1,m\); see, for instance,
\cite{MacWilliamsSloane1977Theory}.  Other complete results include the
classification of Boolean cubic forms in nine variables, which gives the
spectrum of \(\RM(3,9)\) \cite{BrierLangevin2003BooleanCubics}, and recent
algorithmic investigations of small Reed--Muller spectra and weight enumerators
\cite{Jain2024Estimating}.

A particularly active direction concerns high-order families $ \RM(m-c,m),$
where \(c\) is fixed and \(m\) is large.  Carlet and Sol\'e determined the
spectra for \(c=3\) and \(c=4\) \cite{CarletSole2023}.  Carlet then determined
the spectrum of \(\RM(m-5,m)\) for \(m\ge10\), using the
Maiorana--McFarland construction and related structural arguments
\cite{Carlet2024RMSpectrum}.  Lou and Wang determined the spectrum of
\(\RM(m-6,m)\) for \(m\ge12\), and their construction naturally leads to
four-block concatenation methods for the next case
\cite{Lou2025RMm6WeightSpectrum}.  These results naturally motivate the case \(c=7\), 
for which the natural base code is \(\RM(7,14)\).

The present paper addresses the next high-order family \(c=7\) from a
constructive viewpoint.  While a complete determination of
\(W(m-7,m)\) remains open, we establish the following broad
constructive inclusion.
Let
\[
\begin{aligned}
A=\{&
0,128,192,224,240,248,252,254,256,272,288,296,\\
&304,308,312,314,316,318,320,324,328,332\}.
\end{aligned}
\]
Then, for every \(m\ge14\),
\[
        A
        \cup
        \{336+2i \mid 0\le i\le 2^{m-1}-336\}
        \cup
        \{2^m-a \mid a\in A\}
        \subseteq
        W(m-7,m).
\]
Thus the construction gives all even weights in the interval
\([336,2^m-336]\), together with the listed low weights and their complements.

The proof is first carried out in the base code \(\RM(7,14)\).  We use a
four-block concatenation
\[
        g_0\,\|\,g_1\,\|\,g_2\,\|\,(g_1+g_2+g_3),
        \qquad
        g_0,g_3\in\RM(5,12),\quad
        g_1,g_2\in\RM(6,12).
\]
Explicit degree-six monomial constructions provide the required base weights
near the lower end.  These constructions are then combined with known spectrum
inputs for \(\RM(5,10)\) and \(\RM(6,12)\), which supply the shifts needed to
assemble the long interval in \(\RM(7,14)\).  A standard concatenation argument
then lifts the base inclusion to all \(m\ge14\).

In contemporaneous work, Leuenberger and Albrizzio
\cite{LeuenbergerAlbrizzio2026RM714} also investigated the base code
\(\RM(7,14)\) using four-block concatenation and degree-six monomial
constructions. Their concluding remark identifies
\(\{322,326,330,334,354,4378,4380,4382\}\) as eight outstanding
construction targets. The base-code arguments in
Section~\ref{sec:key-section} additionally realize
\(354,4378,4380,4382\). The four low base-code targets not settled
by our constructive inclusion are
\[
        T=\{322,326,330,334\}.
\]

We then carry out a family-wise residual analysis of these targets.
The analysis excludes several natural routes for producing such
weights, including the direct four-block family, a one-point-intersection
minimum-weight branch, punctured affine-flat corrections in a fixed
Carlet-type Maiorana--McFarland shell, selected structured inner
corrections in that shell, and sparse sums of at most eight coordinate
degree-seven monomials.

The main contributions are as follows.
\begin{enumerate}
\item We establish, for every \(m\ge 14\), a constructive inclusion in
   \(W(m-7,m)\) covering all even weights in the full interval
   \([336,2^m-336]\), together with an explicit finite low-weight set
   and the complementary weights.

\item  We prove the base-code inclusion in \(\RM(7,14)\) by combining
   four-block concatenation, explicit degree-six monomial witnesses,
   and known spectrum intervals for \(\RM(5,10)\) and \(\RM(6,12)\).

\item We identify the residual base-code target set
   \(T=\{322,326,330,334\}\) and give exact family-level exclusions
   for several natural construction mechanisms, including four-block
   branches, punctured affine-flat and structured inner corrections
   in a fixed Carlet-type Maiorana--McFarland shell, and sparse
   coordinate degree-seven monomial sums.
\end{enumerate}

The paper is organized as follows. Section~\ref{sec:preliminaries} recalls the
notation and standard Reed--Muller facts used later, including lifting,
concatenation, and low-weight inputs.  Section~\ref{sec:key-section} proves the
constructive spectrum inclusion, first in the base code \(\RM(7,14)\) and then
for all \(m\ge14\).  Section~\ref{sec:residual-section} analyzes the remaining 
target weights in restricted families and delineates the scope of the resulting exclusions.
Section~\ref{sec:conclusion} concludes the paper.  The appendices contain the finite-profile formulations,
affine-geometric arguments, and detailed verification steps used in the
constructive and residual parts.

\section{Preliminaries}
\label{sec:preliminaries}

\subsection{Boolean functions and Reed--Muller codes}

Let \(\mathbb F_2\) be the binary field, and let \(\mathbb F_2^m\) be the
\(m\)-dimensional vector space over \(\mathbb F_2\).  For a positive integer
\(n\), write $ [n]=\{1,\ldots,n\}.$
For a finite set \(X\), write \(\binom{X}{k}\) for the set of all
\(k\)-subsets of \(X\).  Let \(B_m\) denote the set of all Boolean functions from
\(\mathbb F_2^m\) to \(\mathbb F_2\).

Every \(f\in B_m\) has a unique algebraic normal form
\[
        f(x_1,\ldots,x_m)
        =
        \sum_{I\subseteq[m]} a_I\prod_{i\in I}x_i,
        \qquad
        a_I\in\mathbb F_2.
\]
The algebraic degree of \(f\), denoted by \(\deg(f)\), is the largest
\(|I|\) with \(a_I\ne0\).

The \(r\)-th order binary Reed--Muller code of length \(2^m\) is
$ \RM(r,m)=\{f\in B_m \mid \deg(f)\le r\}.$
We identify a Boolean function with its truth table.  Its support and Hamming
weight are $ \supp(f)=\{x\in\mathbb F_2^m \mid f(x)=1\}$ and $\wt(f)=|\supp(f)|.$
The weight spectrum of \(\RM(r,m)\) is $ W(r,m)=\{\wt(f) \mid f\in\RM(r,m)\}.$

An affine \(d\)-flat in \(\mathbb F_2^m\) means a translate \(a+L\), where
\(L\le \mathbb F_2^m\) is a \(d\)-dimensional linear subspace.  For
\(X\subseteq\mathbb F_2^m\), we write \(\mathbf 1_X\) for its indicator
function.

\subsection{Standard Reed--Muller facts}

We shall use the following standard facts.  The minimum nonzero weight of
\(\RM(r,m)\) is $ 2^{m-r}, $
and the minimum-weight codewords are precisely the indicators of affine
\((m-r)\)-flats in \(\mathbb F_2^m\)
\cite{MacWilliamsSloane1977Theory}.  For \(1\le r\le m\), the McEliece divisibility theorem states that every
weight in \(\RM(r,m)\) is divisible by
\(2^{\lfloor (m-1)/r\rfloor}\)
\cite{McEliece1972WeightCongruence}.  Since the all-one function belongs to
\(\RM(r,m)\), the spectrum is symmetric:
\[
        w\in W(r,m)
        \quad\Longleftrightarrow\quad
        2^m-w\in W(r,m).
\]
Invertible affine changes of variables preserve algebraic degree and
Hamming weight.

We also use the classical low-weight classification of Kasami--Tokura and
Kasami--Tokura--Azumi \cite{Kasami1970Weight,Kasami1976WeightEnumeration}.
The specific low-weight consequences needed for the base code \(\RM(7,14)\)
are stated later in Lemma~\ref{lem:low_range}.

\subsection{Lifting and concatenation}

We use two elementary operations repeatedly.  First, if
\(f\in\RM(r,m)\) is viewed as a function of \(m+t\) variables independent of
the last \(t\) variables, then the lifted function belongs to
\(\RM(r,m+t)\) and has weight $ 2^t\wt(f).$
Thus every \(w\in W(r,m)\) yields $ 2^t w\in W(r,m+t).$

Second, we use \(\|\) to denote concatenation.  If \(f_0,f_1\in B_m\), then
\[
        (f_0\|f_1)(x,y)
        =
        (1+y)f_0(x)+y f_1(x),
        \quad
        x\in\mathbb F_2^m,\ y\in\mathbb F_2.
\]
For four blocks indexed by \((y_1,y_2)\), we use the block order
\(00,10,01,11\).  Thus, for
\(f_0,f_1,f_2,f_3\in B_m\),
\[
\begin{aligned}
(f_0\|f_1\|f_2\|f_3)(x,y_1,y_2)
={}&(1+y_1)(1+y_2)f_0(x)
+y_1(1+y_2)f_1(x)\\
&+(1+y_1)y_2f_2(x)
+y_1y_2f_3(x).
\end{aligned}
\]
The degree calculation for the four-block construction used in the main
argument is given in Lemma~\ref{lem:constg}.

\section{A constructive inclusion for the weight spectrum of \(\RM(m-7,m)\)}
\label{sec:key-section}

This section establishes the constructive inclusion. We first treat the
base code \(\RM(7,14)\) and then lift the resulting weight interval to all
\(m\ge 14\). Using a four-block concatenation and explicit degree-\(6\)
monomial constructions, we obtain the required base weights in
\(\RM(7,14)\). These weights are then combined with known
weight-spectrum inclusions for \(\RM(5,10)\) and \(\RM(6,12)\) to form
the desired contiguous interval.

\subsection{The four-block construction and monomial weight formulae}
\label{subsec:four-block-construction}

We first record the concatenation step used throughout the base construction.
The block order is \(00,10,01,11\), as in Section~\ref{sec:preliminaries}.

\begin{lemma}
\label{lem:constg}
Let \(g_0,g_3\in \RM(5,12)\) and \(g_1,g_2\in \RM(6,12)\).  Define
$ g = g_0 \,\|\, g_1 \,\|\, g_2 \,\|\, (g_1+g_2+g_3)$.
Then \(g\in \RM(7,14)\).
\end{lemma}

\begin{proof}
Let \(x_{13}\) and \(x_{14}\) be the two block variables.  Expanding the
concatenation gives
\[
\begin{aligned}
g
&=(1+x_{13})(1+x_{14})g_0
  +x_{13}(1+x_{14})g_1        \\
&\quad +(1+x_{13})x_{14}g_2
  +x_{13}x_{14}(g_1+g_2+g_3).
\end{aligned}
\]
Over \(\mathbb F_2\), the terms \(x_{13}x_{14}g_1\) and
\(x_{13}x_{14}g_2\) cancel.  Hence
\begin{equation}
\label{eq:g-reduced}
        g
        =
        (1+x_{13})(1+x_{14})g_0
        +x_{13}g_1
        +x_{14}g_2
        +x_{13}x_{14}g_3 .
\end{equation}
Since \(\deg(g_0),\deg(g_3)\le 5\) and
\(\deg(g_1),\deg(g_2)\le 6\), every term
in \eqref{eq:g-reduced} has degree at most \(7\).
Thus \(\deg(g)\le 7\), and so \(g\in\RM(7,14)\).

\end{proof}

For later use we record three elementary weight formulae for sums of
coordinate monomials.

\begin{lemma}
\label{eq:2mono}
Let $f = x_{i_{1}} x_{i_{2}} x_{i_{3}} x_{i_{4}} x_{i_{5}} x_{i_{6}} + x_{i_{7}} x_{i_{8}} x_{i_{9}} x_{i_{10}} x_{i_{11}} x_{i_{12}} \in B_{12}$.
Define the index sets as $I = \{i_{1}, i_{2}, i_{3}, i_{4}, i_{5}, i_{6}\}$ and $J = \{i_{7}, i_{8}, i_{9}, i_{10}, i_{11}, i_{12}\}$.
If $|I \cap J| = c$, then the weight of $f$ is given by $\wt(f) = 128 - 2^{c+1}$.
Consequently, the set of possible weights for such functions is
\[
    \{0, 64, 96, 112, 120, 124, 126\}
\]
\end{lemma}
\begin{proof}
Let $f_{1} = x_{i_{1}}x_{i_{2}}x_{i_{3}}x_{i_{4}}x_{i_{5}}x_{i_{6}}$ and $f_{2} = x_{i_{7}}x_{i_{8}}x_{i_{9}}x_{i_{10}}x_{i_{11}}x_{i_{12}}$.
Clearly, for any monomial $m$ of degree $d$ in $B_{12}$, $\wt(m) = 2^{12-d}$.
Since both $f_1$ and $f_2$ have degree 6, we have $\wt(f_{1}) = \wt(f_{2}) = 2^{12-6} = 64$.

By the inclusion-exclusion principle for the weight of Boolean functions: $ \wt(f) = \wt(f_{1}) + \wt(f_{2}) - 2\wt(f_{1}f_{2}) $. For the pairwise product $f_1 f_2$, the set of indices involved is $I \cup J$. The degree of this product is $ |I \cup J| = |I| + |J| - |I \cap J|  = 12 - c $.
Thus, the weight of the product is $\wt(f_{1}f_{2}) = 2^{12-(12-c)} = 2^{c} $.
Substituting these values into the main equation $\wt(f) = 64 + 64 - 2(2^{c}) = 128 - 2^{c+1}$.
This completes the proof.
\end{proof}

\begin{lemma}
\label{eq:3mono}
Let $f = x_{1} x_{2} x_{3} x_{4} x_{5} x_{6} + x_{i_{1}} x_{i_{2}} x_{i_{3}} x_{i_{4}} x_{i_{5}} x_{i_{6}} + x_{i_{7}} x_{i_{8}} x_{i_{9}} x_{i_{10}} x_{i_{11}} x_{i_{12}} \in B_{12}$.
Let $I = \{1, 2, 3, 4, 5, 6\}$, $J = \{i_{1}, i_{2}, i_{3}, i_{4}, i_{5}, i_{6}\}$, and $K = \{i_{7}, i_{8}, i_{9}, i_{10}, i_{11}, i_{12}\}$.
If $|I \cap J| = c_{1}$, $|I \cap K| = c_{2}$, $|J \cap K| = c_{3}$, and $|I \cap J \cap K| = c_{4}$, then
\[
    \wt(f) = 2^{c_{1}+c_{2}+c_{3}-c_{4}-4} - 2^{c_{1}+1} - 2^{c_{2}+1} - 2^{c_{3}+1} + 192.
\]
Consequently, the set of possible weights for such functions is
\[
\{64,96,112,120,124,126,128,136,144,148,152,154,156,160,
162,164,168,172,176\}.
\]

\end{lemma}
\begin{proof}
Let $f_{1} = x_{1}x_{2}x_{3}x_{4}x_{5}x_{6}$, $f_{2} = x_{i_{1}}x_{i_{2}}x_{i_{3}}x_{i_{4}}x_{i_{5}}x_{i_{6}}$, and $f_{3} = x_{i_{7}}x_{i_{8}}x_{i_{9}}x_{i_{10}}x_{i_{11}}x_{i_{12}}$.
Clearly, for any monomial $m$ of degree $d$ in $B_{12}$, $\wt(m) = 2^{12-d}$.
Hence, $\wt(f_{1}) = \wt(f_{2}) = \wt(f_{3}) = 64$.

By the inclusion-exclusion principle for the weight of Boolean functions:
$ \wt(f) = \sum \wt(f_{i}) - 2\sum \wt(f_{i}f_{j}) + 4\wt(f_{1}f_{2}f_{3}) $.
For the pairwise products, the degree is $|I \cup J| = 12 - c_{1}$. Thus,
$\wt(f_{1}f_{2}) = 2^{12-(12-c_{1})} = 2^{c_{1}}$.
Similarly, $\wt(f_{1}f_{3}) = 2^{c_{2}}$ and $\wt(f_{2}f_{3}) = 2^{c_{3}}$.

Moreover, for the triple product, the degree is
$|I \cup J \cup K| = 18 - c_{1} - c_{2} - c_{3} + c_{4}$.
Therefore,
$\wt(f_{1}f_{2}f_{3}) = 2^{12 - (18 - c_{1} - c_{2} - c_{3} + c_{4})} = 2^{c_{1}+c_{2}+c_{3}-c_{4}-6}$.

Substituting these into the main equation:
\begin{align*}
    \wt(f) &= (64+64+64) - 2(2^{c_{1}} + 2^{c_{2}} + 2^{c_{3}}) + 4(2^{c_{1}+c_{2}+c_{3}-c_{4}-6}) \\
    &= 2^{c_{1}+c_{2}+c_{3}-c_{4}-4} - 2^{c_{1}+1} - 2^{c_{2}+1} - 2^{c_{3}+1} + 192.
\end{align*}
This completes the proof.
\end{proof}

\begin{lemma}
\label{eq:4mono}
Let \( f = m_1 + m_2 + m_3 + m_4 \) be a Boolean function in \( B_{12} \), where each \( m_k \) is a monomial of degree 6 with index set \( I_k \subset \{1, \dots, 12\} \). For \(1 \le i < j \le 4\) and \(1 \le i < j < k \le 4\), define $c_{ij} = |I_i \cap I_j|, t_{ijk} = |I_i \cap I_j \cap I_k| , q = |I_1 \cap I_2 \cap I_3 \cap I_4|$, then
\begin{align*}
\wt(f) = & 256 - 2\sum_{1 \le i < j \le 4} 2^{c_{ij}} + 4\sum_{1 \le i < j < k \le 4} 2^{(c_{ij} + c_{ik} + c_{jk}) - t_{ijk} - 6} - 8 \cdot 2^{(\sum c_{ij}) - (\sum t_{ijk}) + q - 12}.
\end{align*}
Consequently, the set of possible weights for such functions is
\[
\begin{aligned}
\{&0,64,96,112,120,124,126,128,136,144,148,152,154,156,158,160,162,164,166,168,172,\\
&174,176,180,182,184,186,188,190,192,194,196,198,200,202,204,208,210,212,216,220\}.
\end{aligned}
\]
\end{lemma}

\begin{proof}
Let $A_i = \{x\in\mathbb{F}_2^{12} \mid m_i(x)=1\}$ be the support of $m_i$. The weight of $f$ corresponds to the size of the symmetric difference of the sets $A_i$:
\begin{equation}
\label{eq:symdiff4}
\wt(f) = \left|\Delta_{i=1}^4 A_i\right| = \sum |A_i| - 2\sum |A_i\cap A_j| + 4\sum |A_i\cap A_j\cap A_k| - 8|\cap_{i=1}^4 A_i|.
\end{equation}
For any subset of indices $S \subseteq \{1,2,3,4\}$, the intersection $\bigcap_{i\in S} A_i$ corresponds to the support of the monomial $\prod_{i\in S} m_i$, which has degree $|\bigcup_{i\in S} I_i|$. Thus, $|\bigcap_{i\in S} A_i| = 2^{12 - |\bigcup_{i\in S} I_i|}$.

Clearly, $|A_i| = 2^{12-6} = 64$, so $\sum |A_i| = 256$.
For pairwise intersections, $|\bigcup_{i<j} I_i| = 12 - c_{ij}$, yielding $|A_i \cap A_j| = 2^{c_{ij}}$.
For triple intersections, by the inclusion-exclusion principle, $|\bigcup_{i<j<k} I_i| = 18 - (c_{ij}+c_{ik}+c_{jk}) + t_{ijk}$, implying $|A_i \cap A_j \cap A_k| = 2^{c_{ij}+c_{ik}+c_{jk}-t_{ijk}-6}$.
Similarly, for the quadruple intersection, $|\bigcup_{i=1}^4 I_i| = 24 - \sum c_{ij} + \sum t_{ijk} - q$, which gives the final term's exponent as $(\sum c_{ij}) - (\sum t_{ijk}) + q - 12$.

Substituting these terms into \eqref{eq:symdiff4} yields the result.
\end{proof}

\subsection{Base weights in \texorpdfstring{\(\RM(7,14)\)}{RM(7,14)}}
\label{subsec:certified-base-weights}

We now specialize the construction to the base code \(\RM(7,14)\).  The
classical low-weight results first give the complete list of weights below
\(2.5\cdot 2^7=320\).

\begin{lemma}[\cite{Kasami1970Weight, Kasami1976WeightEnumeration}]
\label{lem:low_range}
The set of all weights $w$ in the range $ w<2.5\cdot 2^7=320 $ for the Reed--Muller code $\RM(7,14)$ is
\[
\{0,128,192,224,240,248,252,254,
256,272,288,296,304,308,312,314,316,318\}.
\]
\end{lemma}

\begin{construction}
\label{con:add_2mono}
Let $g_1, g_2 \in B_{12}$ be Boolean functions defined by
\[
\begin{aligned}
g_1 &= x_1 x_2 x_3 x_4 x_5 x_6 + x_{i_1} x_{i_2} x_{i_3} x_{i_4} x_{i_5} x_{i_6} + x_{i_7} x_{i_8} x_{i_9} x_{i_{10}} x_{i_{11}} x_{i_{12}}, \\
g_2 &= x_1 x_2 x_3 x_4 x_5 x_6 + x_{i_1} x_{i_2} x_{i_3} x_{i_4} x_{i_5} x_{i_6} + x_{i_{13}} x_{i_{14}} x_{i_{15}} x_{i_{16}} x_{i_{17}} x_{i_{18}},
\end{aligned}
\]
where each displayed product is a squarefree monomial of degree \(6\). Then construct the function $g$ as
$ g = 0 \,\|\, g_1 \,\|\, g_2 \,\|\, (g_1+g_2) $.
\end{construction}

For \(I=\{i_1,\ldots,i_6\}\subseteq[12]\), write $ m_I=\prod_{i\in I}x_i .$
When no confusion can arise, we write \(m_{i_1,\ldots,i_6}\) for
\(m_{\{i_1,\ldots,i_6\}}\).  All sums of Boolean functions are taken over
\(\mathbb F_2\).

In the explicit witnesses below, displayed monomial lists may contain
repetitions unless pairwise distinctness is explicitly required.  Equal
displayed monomials cancel in pairs over \(\mathbb F_2\); for instance,
\(m_0+m_0+m_I=m_I\).  Thus the stated construction families include the
lower-term specializations obtained after such cancellations.

The next propositions supply the additional base weights required to
connect the low-weight range to the long intervals obtained by lifting.

\begin{proposition}
\label{prop:318446}
Boolean functions obtained from Construction~\ref{con:add_2mono} attain
every integer congruent to \(2 \pmod 4\) in
\([314,318]\cup[446,450]\).
\end{proposition}

\begin{proof}
Put \(m_0=m_{1,2,3,4,5,6}\).  For
\(G=0\|g_1\|g_2\|(g_1+g_2)\), one has
$ \wt(G)=\wt(g_1)+\wt(g_2)+\wt(g_1+g_2). $
The relevant values in \([314,318]\cup[446,450]\) are
\(314,318,446,450\).  The following table gives explicit witnesses.
The three weights in the last column are computed from
Lemmas~\ref{eq:2mono} and~\ref{eq:3mono}.

\begin{center}
\scriptsize
\setlength{\tabcolsep}{2pt}
\renewcommand{\arraystretch}{1.15}
\begin{tabularx}{\linewidth}{@{}cYYc@{}}
\toprule
\(w\) & \(g_1\) & \(g_2\) &
\((\wt(g_1),\wt(g_2),\wt(g_1+g_2))\) \\
\midrule
\(314\)
& \(m_0\)
& \(m_0+\biwmon{7,8,9,10,11,12}+\biwmon{1,2,3,4,5,7}\)
& \((64,126,124)\) \\
\addlinespace
\(318\)
& \(m_0+\biwmon{1,2,3,4,5,7}+\biwmon{1,2,3,4,5,8}\)
& \(m_0+\biwmon{1,2,3,4,5,7}+\biwmon{6,7,9,10,11,12}\)
& \((64,128,126)\) \\
\addlinespace
\(446\)
& \(m_0+\biwmon{1,2,7,8,9,10}+\biwmon{1,3,4,7,8,11}\)
& \(m_0+\biwmon{1,2,7,8,9,10}+\biwmon{2,5,6,9,10,12}\)
& \((160,160,126)\) \\
\addlinespace
\(450\)
& \(m_0+\biwmon{1,2,3,7,8,9}+\biwmon{1,2,3,10,11,12}\)
& \(m_0+\biwmon{1,2,3,7,8,9}+\biwmon{4,5,6,7,8,9}\)
& \((148,176,126)\) \\
\bottomrule
\end{tabularx}
\end{center}

For each row, the sum of the three weights in the last column is the
target value \(w\).  Hence Construction~\ref{con:add_2mono} attains all
integers congruent to \(2\pmod 4\) in the stated intervals.
\end{proof}

\begin{proposition}
\label{prop:bit_flipping}
Let $g_1, g_2 \in B_{12}$ be Boolean functions on $12$ variables. Consider the function $G \in B_{14}$ constructed by the concatenation $ G = 0 \,\|\, g_1 \,\|\, g_2 \,\|\, (g_1 + g_2)$. 
Let $\mathbf{1}$ denote the all-one function in $B_{12}$. The weights of the functions obtained by flipping the bits of specific components of $G$ are given by:
\begin{align*}
\wt\bigl(0\,\|\,(g_1+\mathbf{1})\,\|\,g_2\,\|\,(g_1+g_2)\bigr)
&=\wt(G)+4096-2\wt(g_1),\\
\wt\bigl(0\,\|\,g_1\,\|\,(g_2+\mathbf{1})\,\|\,(g_1+g_2)\bigr)
&=\wt(G)+4096-2\wt(g_2),\\
\wt\bigl(0\,\|\,g_1\,\|\,g_2\,\|\,(g_1+g_2+\mathbf{1})\bigr)
&=\wt(G)+4096-2\wt(g_1+g_2).
\end{align*}
\end{proposition}

\begin{proof}
Replacing \(g_1\) by \(g_1+\mathbf{1}\) complements a block of length \(2^{12}=4096\),
so its weight changes from \(\wt(g_1)\) to \(4096-\wt(g_1)\).  Hence
$ \wt\bigl(0\|(g_1+\mathbf{1})\|g_2\|(g_1+g_2)\bigr)
=\wt(G)+4096-2\wt(g_1).$
The other two identities follow identically by complementing the
corresponding block.
\end{proof}

\begin{proposition}
\label{prop:409842384354}
Let \(g_1,g_2\in B_{12}\) be defined by
\[
\begin{aligned}
g_1 &=
m_0+m_1+m_2,\\
g_2 &=
m_0+m_1+h,
\end{aligned}
\]
where \(m_0=m_{1,2,3,4,5,6}\), and \(m_1,m_2,h\) are squarefree
monomials of degree \(6\), with repetitions allowed in the displayed
form and cancelled over \(\mathbb F_2\).  For suitable choices of the
monomials and \(a_1,a_2,a_3\in\mathbb F_2\), Boolean functions of the
form $ 0\,\|\,(g_1+a_1)\,\|\,(g_2+a_2)\,\|\,(g_1+g_2+a_3)$
attain every integer congruent to \(2 \pmod 4\) in $ [4098,4102]\cup[4238,4250]\cup[4354,4362].$
\end{proposition}

\begin{proof}
For \(G=0\|g_1\|g_2\|(g_1+g_2)\), put
$ \Omega(g_1,g_2)=\bigl(\wt(g_1),\wt(g_2),\wt(g_1+g_2)\bigr).$
The following two tables give explicit witnesses.  The column ``flip''
indicates which of the three nonzero blocks is complemented.  The
listed triples \(\Omega(g_1,g_2)\) are computed from
Lemmas~\ref{eq:2mono} and~\ref{eq:3mono}.

\begin{center}
\scriptsize
\setlength{\tabcolsep}{2pt}
\renewcommand{\arraystretch}{1.15}
\begin{tabularx}{\linewidth}{@{}cYYcc@{}}
\toprule
\(w\) & \(g_1\) & \(g_2\) &
\(\Omega(g_1,g_2)\) & flip \\
\midrule
\(4098\)
& \(m_0+\biwmon{7,8,9,10,11,12}+\biwmon{1,7,8,9,10,11}\)
& \(m_0\)
& \((126,64,64)\)
& \(g_1\) \\
\addlinespace
\(4102\)
& \(m_0+\biwmon{7,8,9,10,11,12}+\biwmon{1,2,7,8,9,10}\)
& \(m_0\)
& \((154,64,96)\)
& \(g_1\) \\
\addlinespace
\(4238\)
& \(m_0+\biwmon{1,2,3,7,8,9}+\biwmon{1,4,5,6,10,11}\)
& \(m_0+\biwmon{1,2,3,7,8,9}+\biwmon{2,3,7,8,9,12}\)
& \((148,120,126)\)
& \(g_1+g_2\) \\
\addlinespace
\(4242\)
& \(m_0+\biwmon{1,2,7,8,9,10}+\biwmon{1,2,3,7,8,9}\)
& \(m_0+\biwmon{1,2,7,8,9,10}+\biwmon{7,8,9,10,11,12}\)
& \((120,154,112)\)
& \(g_1\) \\
\addlinespace
\(4246\)
& \(m_0+\biwmon{1,2,3,7,8,9}+\biwmon{4,5,6,7,10,11}\)
& \(m_0+\biwmon{1,2,3,7,8,9}+\biwmon{1,2,3,8,9,12}\)
& \((164,112,126)\)
& \(g_1+g_2\) \\
\addlinespace
\(4250\)
& \(m_0+\biwmon{1,2,3,4,5,7}+\biwmon{1,2,3,4,7,8}\)
& \(m_0+\biwmon{1,2,3,4,5,7}+\biwmon{5,6,9,10,11,12}\)
& \((96,124,126)\)
& \(g_1\) \\
\bottomrule
\end{tabularx}
\end{center}

\begin{center}
\scriptsize
\setlength{\tabcolsep}{2pt}
\renewcommand{\arraystretch}{1.15}
\begin{tabularx}{\linewidth}{@{}cYYcc@{}}
\toprule
\(w\) & \(g_1\) & \(g_2\) &
\(\Omega(g_1,g_2)\) & flip \\
\midrule
\(4354\)
& \(m_0+\biwmon{1,2,3,7,8,9}+\biwmon{1,7,8,10,11,12}\)
& \(m_0+\biwmon{1,2,3,7,8,9}+\biwmon{7,8,9,10,11,12}\)
& \((160,162,64)\)
& \(g_1+g_2\) \\
\addlinespace
\(4358\)
& \(m_0+\biwmon{1,2,3,7,8,9}+\biwmon{7,8,9,10,11,12}\)
& \(m_0+\biwmon{1,2,3,7,8,9}+\biwmon{4,7,8,9,10,11}\)
& \((162,164,64)\)
& \(g_1+g_2\) \\
\addlinespace
\(4362\)
& \(m_0+\biwmon{1,2,3,7,8,9}+\biwmon{4,5,7,10,11,12}\)
& \(m_0+\biwmon{1,2,3,7,8,9}+\biwmon{4,5,6,10,11,12}\)
& \((168,162,64)\)
& \(g_1+g_2\) \\
\bottomrule
\end{tabularx}
\end{center}

For each row, let \(\Omega(g_1,g_2)=(\omega_1,\omega_2,\omega_{12})\).
Then $ \wt(G)=\omega_1+\omega_2+\omega_{12}.$
If the flipped block is \(g_1\), \(g_2\), or \(g_1+g_2\), respectively,
Proposition~\ref{prop:bit_flipping} gives the final weight as
\[
        \wt(G)+4096-2\omega_1,\qquad
        \wt(G)+4096-2\omega_2,\qquad
        \wt(G)+4096-2\omega_{12}.
\]
Applying the indicated flip in each row gives precisely the target
weight \(w\).  Therefore all weights congruent to \(2\pmod 4\) in the
three stated intervals are attained.
\end{proof}

To obtain the remaining high base weights, we also allow the sum component to contain four degree-six monomials.

\begin{construction}
\label{con:add_4mono}
Let $g_1, g_2 \in B_{12}$ be Boolean functions defined by
\[
\begin{aligned}
g_1 &= x_1 x_2 x_3 x_4 x_5 x_6 + x_{i_1} x_{i_2} x_{i_3} x_{i_4} x_{i_5} x_{i_6} + x_{i_7} x_{i_8} x_{i_9} x_{i_{10}} x_{i_{11}} x_{i_{12}}, \\
g_2 &= x_1 x_2 x_3 x_4 x_5 x_6 + x_{j_1} x_{j_2} x_{j_3} x_{j_4} x_{j_5} x_{j_6} + x_{j_7} x_{j_8} x_{j_9} x_{j_{10}} x_{j_{11}} x_{j_{12}},
\end{aligned}
\]
where each displayed product is a squarefree monomial of degree \(6\). We then construct the function $g$ as
$g = 0 \,\|\, g_1 \,\|\, g_2 \,\|\, (g_1+g_2)$.

\end{construction}

\begin{proposition}
\label{prop:4380}
Let \(m_0=m_{1,2,3,4,5,6}\).  Let \(g_1,g_2\in B_{12}\) be defined by
\[
        g_1=m_0+m_1+m_2,\qquad
        g_2=m_0+m_3+m_4,
\]
where \(m_1,m_2,m_3,m_4\) are pairwise distinct squarefree monomials of
degree \(6\), not necessarily distinct from the fixed monomial \(m_0\).
For suitable choices of \(m_1,m_2,m_3,m_4\) and
\(a_1,a_2,a_3\in\mathbb F_2\), Boolean functions of the form
$0\,\|\,(g_1+a_1)\,\|\,(g_2+a_2)\,\|\,(g_1+g_2+a_3) $
attain every integer congruent to \(0 \pmod 4\) in \([4380,4384]\).
\end{proposition}

\begin{proof}
For \(G=0\|g_1\|g_2\|(g_1+g_2)\), the following table gives witnesses.
In both rows the \(g_1\)-block is complemented.

\begin{center}
\scriptsize
\setlength{\tabcolsep}{2pt}
\renewcommand{\arraystretch}{1.15}
\begin{tabularx}{\linewidth}{@{}cYYc@{}}
\toprule
\(w\) & \(g_1\) & \(g_2\) &
\((\wt(g_1),\wt(g_2),\wt(g_1+g_2))\) \\
\midrule
\(4380\)
& \(m_0+m_0+\biwmon{1,2,3,4,5,7}\)
& \(m_0+\biwmon{1,2,7,8,9,10}+\biwmon{3,6,7,8,11,12}\)
& \((64,172,176)\) \\
\addlinespace
\(4384\)
& \(m_0+m_0+\biwmon{1,2,3,4,7,8}\)
& \(m_0+\biwmon{1,2,3,9,10,11}+\biwmon{7,8,9,10,11,12}\)
& \((64,162,190)\) \\
\bottomrule
\end{tabularx}
\end{center}

The displayed \(g_1\)-terms use one intentional cancellation against
\(m_0\).  The three weights in the last column are computed from
Lemmas~\ref{eq:2mono}--\ref{eq:4mono}.  For the first row,
$ \wt(G)=64+172+176=412, $ then we have $ 412+4096-2\cdot64=4380.$
For the second row,
$ \wt(G)=64+162+190=416,$ then we have $ 416+4096-2\cdot64=4384.$
The conclusion follows from Proposition~\ref{prop:bit_flipping}.
\end{proof}

\begin{proposition}
\label{prop:43824390}
Let \(m_0=m_{1,2,3,4,5,6}\).  Let \(g_1,g_2\in B_{12}\) be defined by
\[
        g_1=m_0+m_1+m_2,\qquad
        g_2=m_0+m_3+m_4,
\]
where \(m_1,m_2,m_3,m_4\) are pairwise distinct squarefree monomials of
degree \(6\), not necessarily distinct from the fixed monomial \(m_0\).
For suitable choices of \(m_1,m_2,m_3,m_4\) and
\(a_1,a_2,a_3\in\mathbb F_2\), Boolean functions of the form
$ 0\,\|\,(g_1+a_1)\,\|\,(g_2+a_2)\,\|\,(g_1+g_2+a_3) $
attain every integer congruent to \(2 \pmod 4\) in
\(\{4382\}\cup[4390,4394]\).
\end{proposition}

\begin{proof}
For \(G=0\|g_1\|g_2\|(g_1+g_2)\), the following table gives witnesses.
In all rows the \(g_1\)-block is complemented.

\begin{center}
\scriptsize
\setlength{\tabcolsep}{2pt}
\renewcommand{\arraystretch}{1.15}
\begin{tabularx}{\linewidth}{@{}cYYc@{}}
\toprule
\(w\) & \(g_1\) & \(g_2\) &
\((\wt(g_1),\wt(g_2),\wt(g_1+g_2))\) \\
\midrule
\(4382\)
& \(m_0+m_0+\biwmon{1,2,3,4,7,8}\)
& \(m_0+\biwmon{1,2,3,7,9,10}+\biwmon{5,6,9,10,11,12}\)
& \((64,168,182)\) \\
\addlinespace
\(4390\)
& \(m_0+m_0+\biwmon{1,2,3,4,7,8}\)
& \(m_0+\biwmon{1,2,7,8,9,10}+\biwmon{5,6,9,10,11,12}\)
& \((64,172,186)\) \\
\addlinespace
\(4394\)
& \(m_0+m_0+\biwmon{3,5,6,9,11,12}\)
& \(m_0+\biwmon{1,2,4,7,8,10}+\biwmon{3,7,8,9,10,11}\)
& \((64,164,198)\) \\
\bottomrule
\end{tabularx}
\end{center}

Again, each displayed \(g_1\) uses one intentional cancellation against
\(m_0\).  The three weights in the last column are computed from
Lemmas~\ref{eq:2mono}--\ref{eq:4mono}.  The corresponding unflipped
weights are $ 64+168+182=414,64+172+186=422$ and $
64+164+198=426.$
Complementing the \(g_1\)-block and applying
Proposition~\ref{prop:bit_flipping} gives
$ 414+4096-2\cdot64=4382,
422+4096-2\cdot64=4390 $ and $
426+4096-2\cdot64=4394. $
Hence all stated weights are attained.
\end{proof}

\begin{construction}
\label{con:add_6mono}
Let \(g_1,g_2\in B_{12}\) be Boolean functions defined by
\[
\begin{aligned}
g_1 &=
x_{i_1}x_{i_2}x_{i_3}x_{i_4}x_{i_5}x_{i_6}
+x_{i_7}x_{i_8}x_{i_9}x_{i_{10}}x_{i_{11}}x_{i_{12}}
+x_{i_{13}}x_{i_{14}}x_{i_{15}}x_{i_{16}}x_{i_{17}}x_{i_{18}},\\
g_2 &=
x_{j_1}x_{j_2}x_{j_3}x_{j_4}x_{j_5}x_{j_6}
+x_{j_7}x_{j_8}x_{j_9}x_{j_{10}}x_{j_{11}}x_{j_{12}}
+x_{j_{13}}x_{j_{14}}x_{j_{15}}x_{j_{16}}x_{j_{17}}x_{j_{18}},
\end{aligned}
\]
where each group of six indices consists of distinct elements of
\(\{1,\ldots,12\}\), and the six resulting monomials are pairwise
distinct. We then construct the function \(g\) as
$g=0\,\|\,g_1\,\|\,g_2\,\|\,(g_1+g_2).$
\end{construction}

\begin{proposition}
\label{prop:base_interval}
Boolean functions obtained from
Constructions~\ref{con:add_2mono},~\ref{con:add_4mono} and~\ref{con:add_6mono}
collectively attain every integer congruent to \(2 \pmod 4\) in
\([338,474]\cup[538,542]\).
\end{proposition}

\begin{proof}
Let \(m_0=m_{1,2,3,4,5,6}\).  For
\(G=0\|g_1\|g_2\|(g_1+g_2)\), we use
\[
        \wt(G)=\wt(g_1)+\wt(g_2)+\wt(g_1+g_2).
\]
The five endpoint and exceptional witnesses \(338,354,474,538,542\)
are displayed below.  The remaining target weights congruent to \(2\pmod4\) in \([338,474]\) are
recorded in Table~\ref{tab:base-interval-witnesses}
(Appendix~\ref{app:remaining-monomial-witnesses}).
Together, the displayed
witnesses and Table~\ref{tab:base-interval-witnesses} cover exactly all
integers congruent to \(2\pmod4\) in \([338,474]\cup[538,542]\).

\noindent\textbf{Weight \(338\).}
Let $m_1=m_{7,8,9,10,11,12},$ and $m_2=m_{1,2,3,4,7,8}.$
Set \(g_1=m_0+m_1+m_1=m_0\) and
\(g_2=m_0+m_1+m_2\).  Then $ \bigl(\wt(g_1),\wt(g_2),\wt(g_1+g_2)\bigr)=(64,154,120),$
and hence \(\wt(G)=64+154+120=338\).

\noindent\textbf{Weight \(354\).}
Let
\[
\begin{aligned}
m_1&=m_{3,4,7,8,11,12},&
m_2&=m_{3,4,6,7,8,11},&
m_3&=m_{1,3,4,7,8,11},\\
m_4&=m_{1,4,5,8,9,11},&
m_5&=m_{2,5,6,9,10,12},&
m_6&=m_{2,3,5,6,9,12}.
\end{aligned}
\]
Set \(g_1=m_1+m_2+m_3\) and \(g_2=m_4+m_5+m_6\).  Then
$\bigl(\wt(g_1),\wt(g_2),\wt(g_1+g_2)\bigr)
=(64,120,170),$
and hence \(\wt(G)=64+120+170=354\).  The detailed verification of
these weight calculations is given in Appendix~\ref{app:354}.

\noindent\textbf{Weight \(474\).}
Let
\[
\begin{aligned}
m_1&=m_{1,2,3,4,6,8},&
m_2&=m_{1,4,7,9,10,12},\\
m_3&=m_{2,5,6,7,10,12},&
m_4&=m_{7,8,9,10,11,12}.
\end{aligned}
\]
Set \(g_1=m_0+m_1+m_2\) and \(g_2=m_0+m_3+m_4\).  Then
$ \bigl(\wt(g_1),\wt(g_2),\wt(g_1+g_2)\bigr)
=(120,162,192), $
and hence \(\wt(G)=120+162+192=474\).

\noindent\textbf{Weight \(538\).}
Let
\[
\begin{aligned}
m_1&=m_{4,5,6,7,10,11},&
m_2&=m_{1,2,3,7,8,11},\\
m_3&=m_{7,8,9,10,11,12},&
m_4&=m_{1,3,5,9,10,12}.
\end{aligned}
\]
Set \(g_1=m_0+m_1+m_2\) and \(g_2=m_0+m_3+m_4\).  Then
$\bigl(\wt(g_1),\wt(g_2),\wt(g_1+g_2)\bigr)
=(168,162,208),$
and hence \(\wt(G)=168+162+208=538\).

\noindent\textbf{Weight \(542\).}
Let
\[
\begin{aligned}
m_1&=m_{1,4,9,10,11,12},&
m_2&=m_{3,5,6,10,11,12},\\
m_3&=m_{3,5,6,7,8,9},&
m_4&=m_{1,2,4,7,8,9}.
\end{aligned}
\]
Set \(g_1=m_0+m_1+m_2\) and \(g_2=m_0+m_3+m_4\).  Then
$ \bigl(\wt(g_1),\wt(g_2),\wt(g_1+g_2)\bigr)
=(168,176,198), $
and hence \(\wt(G)=168+176+198=542\).

For each row of Table~\ref{tab:base-interval-witnesses}, the same
identity \(\wt(G)=\wt(g_1)+\wt(g_2)+\wt(g_1+g_2)\) is used, and the
three component weights are listed explicitly in the table.  This proves
the proposition.
\end{proof}

\subsection{Base interval completion and lifting}

To complete the base case, we combine the explicit constructions established
above with the following known weight-spectrum inclusions and then apply
complementation symmetry.

\begin{lemma}[\cite{Carlet2024RMSpectrum}]
\label{lem:RM5_10_bounds}
The set of weights in $\RM(5, 10)$ contains all even integers lying in the interval $[72, 952]$.
\end{lemma}

\begin{corollary}
\label{cor:RM5_12_bounds}
The weight spectrum of \(\RM(5,12)\) contains every integer congruent
to \(0 \pmod 8\) in the interval $[288,3808].$
\end{corollary}

\begin{proof}
For \(f\in\RM(5,10)\), define $\widetilde f(x_1,\ldots,x_{12})=f(x_1,\ldots,x_{10}).$
Then \(\widetilde f\in\RM(5,12)\) and $\wt(\widetilde f)=4\wt(f).$
The result therefore follows from Lemma~\ref{lem:RM5_10_bounds}.
\end{proof}

\begin{lemma}[\cite{Lou2025RMm6WeightSpectrum}]
\label{lem:RM6_12_bounds}
The set of weights in $\RM(6, 12)$ contains all even integers lying in the interval $[152, 3944]$.
\end{lemma}

\begin{proposition}
\label{prop:endconclusion}
Let $A = \{0, 128, 192, 224, 240, 248, 252, 254, 256, 272, 288, 296, 304, 308, 312,\\
314, 316, 318, 320, 324, 328, 332\}$ and $S$ be the set of all weights in $\RM(7, 14)$. Then
\[
A \cup \{336 + 2i\} \cup \{2^{14} - a \mid a \in A\} \subseteq S,
\]
where $i$ ranges over the set of consecutive integers from $0$ to $2^{13} - 336$.
\end{proposition}

\begin{proof}
Consider the functions of the form $g = g_0 \| g_1 \| g_2 \| (g_1 + g_2 + g_3)$, where $g_0, g_3 \in \RM(5, 12)$ and $g_1, g_2 \in \RM(6, 12)$. By Lemma \ref{lem:constg}, $g \in \RM(7, 14)$. Clearly, $\wt(g) = \wt(g_0) + \wt(0 \| g_1 \| g_2 \| (g_1 + g_2 + g_3)).$

We first establish a contiguous range of weights congruent to
\(2\pmod 4\). By Corollary~\ref{cor:RM5_12_bounds}, the first block
\(g_0\in\RM(5,12)\) can attain every weight
\(t\in8\mathbb{Z}\cap[288,3808]\).
Adding such a block to the constructions in
Propositions~\ref{prop:318446}, \ref{prop:409842384354},
and~\ref{prop:43824390} shifts each base weight \(w\) to \(w+t\).
The base intervals \([314,318]\) and \([446,450]\) therefore yield
every weight congruent to \(2\pmod 4\) in \([602,4258]\).
Similarly, the base intervals \([4098,4102]\), \([4354,4362]\), and
\([4390,4394]\), together with the isolated base weight \(4382\),
yield every such weight in \([4382,8190]\).
Thus, at this stage, the covered range is $[602,4258]\cup[4382,8190].$

We next fill the gap between these two intervals. Shifting the base
weights congruent to \(2\pmod 4\) in \([538,542]\), provided by
Proposition~\ref{prop:base_interval}, by every
\(t\in8\mathbb{Z}\cap[288,3808]\) yields all such weights in
$[826,4350]=[538+288,\,542+3808].$
Together with the previously obtained interval \([602,4258]\) and the
weights in \([4354,4362]\) provided by
Proposition~\ref{prop:409842384354}, this extends the covered range to
$[602,4362]\cup[4382,8190].$

It remains to extend the lower endpoint and fill the gap between \(4362\)
and \(4382\). For the lower endpoint, choose
\(g_0\in\RM(5,12)\) with \(\wt(g_0)=128\). Shifting the base interval
\([338,474]\) from Proposition~\ref{prop:base_interval} by this fixed
weight yields every integer congruent to \(2\pmod 4\) in $ [466,602]=[338+128,\,474+128].$
Together with the unshifted interval \([338,474]\) and the previously
covered range, this gives every weight congruent to \(2\pmod 4\) in
$[338,4362]\cup[4382,8190].$

Finally, Proposition~\ref{prop:409842384354} provides every weight
congruent to \(2\pmod 4\) in \([4238,4250]\). Applying the same fixed
shift of \(128\) yields $\{4366,4370,4374,4378\},$
which are precisely the missing weights between \(4362\) and \(4382\).
Together with the already attained weight \(4382\), these weights connect
the two ranges. Consequently, the weight spectrum contains every integer
\(w\equiv2\pmod4\) in $[338,8190].$

Next, we consider the weights congruent to \(0\pmod 4\). For any
\(f\in\RM(6,12)\), the concatenated function $ g'=0\,\|\,0\,\|\,f\,\|\,f $
belongs to \(\RM(7,14)\) and satisfies $\wt(g')=2\wt(f).$
By Lemma~\ref{lem:RM6_12_bounds}, \(\RM(6,12)\) attains every even
weight in \([152,3944]\). Hence, \(\RM(7,14)\) attains every integer
congruent to \(0\pmod 4\) in $[304,7888].$

To extend the upper endpoint, we combine Proposition~\ref{prop:4380}
with Corollary~\ref{cor:RM5_12_bounds}. Shifting the base weights
\(4380\) and \(4384\) by every
\(t\in8\mathbb{Z}\cap[288,3808]\) yields every integer congruent to
\(0\pmod 4\) in $[4668,8192].$
Since this interval overlaps \([304,7888]\), we conclude that every
integer congruent to \(0\pmod 4\) in $ [304,8192] $ is attained.

Combining this result with the weights congruent to \(2\pmod 4\)
established above, we obtain every even weight in \([336,8192]\).
Moreover, since the all-one function belongs to \(\RM(7,14)\), the
weight spectrum \(S\) is invariant under complementation, that is,
\(w\in S\) implies \(2^{14}-w\in S\). Therefore,
\[
A\cup
\{336+2i \mid 0\le i\le 2^{13}-336\}
\cup
\{2^{14}-a \mid a\in A\}
\subseteq S.
\]
\end{proof}

\begin{theorem}
\label{thm:main-thm}
Let $A = \{0, 128, 192, 224, 240, 248, 252, 254, 256, 272, 288, 296, 304, 308, 312,\\
314, 316, 318, 320, 324, 328, 332\}$ and $S$ be the set of all weights in $\RM(m-7, m)$, where $m \ge 14 $. Then
\[
 A \cup \{336 + 2i\} \cup \{2^{m} - a \mid a \in A\} \subseteq S ,
\]
where $i$ ranges over the set of consecutive integers from $0$ to $2^{m-1} - 336$.
\end{theorem}

\begin{proof}
By Proposition \ref{prop:endconclusion}, the result is correct for $m = 14$. Assuming the result is correct for $m$, we now prove that it is also correct for $m+1$. Let $S$ be the set of all weights in $\RM(m-6, m+1)$. Since $0\|f \in \RM(m-6, m+1)$ for any $f \in \RM(m-7, m)$, and $\wt(0\|f) = \wt(f)$, we have
\begin{equation}
\label{eq:induction-first-range}
A\cup
\{336+2i\mid 0\le i\le 2^{m-1}-336\}
\subseteq S.
\end{equation}

Let $g_1 \in \RM(m-7, m)$ with $\wt(g_1) = 336$. Since $g_1\|g_2 \in \RM(m-6, m+1)$ for any $g_2 \in \RM(m-7, m)$ and $\wt(g_1\|g_2) = 336 + \wt(g_2)$, we have
\begin{equation}
\label{eq:induction-second-range}
\{336+2i\mid 2^{m-1}-336\le i\le 2^{m-1}-168\}
\subseteq S.
\end{equation}

Moreover, $\wt(f + 1) = 2^{m+1} - \wt(f)$, for any $f \in \RM(m-6, m+1)$. Then, by
\eqref{eq:induction-first-range} and
\eqref{eq:induction-second-range}, we have
\[
A \cup \{336 + 2i\} \cup \{2^{m+1} - a \mid a \in A\} \subseteq S,
\]

where $i$ ranges over the set of consecutive integers from $0$ to $2^m - 336$.
\end{proof}

\section{Analysis of the remaining target weights in \texorpdfstring{\(\RM(7,14)\)}{RM(7,14)}}
\label{sec:residual-section}

Let \(T=\{322,326,330,334\}\).  After the constructive inclusion of
Section~\ref{sec:key-section}, these are precisely the low base-code
targets in \(\RM(7,14)\) not settled by the construction, up to
complementation.  By spectrum symmetry, it suffices to analyze the
weights in \(T\).

This section gives a family-wise residual analysis of these four
weights.  Each result below is asserted only for the explicitly stated
family, normal form, and parameter range.  Within this convention, we
rule out several natural mechanisms for producing the residual targets:
two four-block branches, punctured affine-flat corrections in a fixed
Carlet-type Maiorana--McFarland outer shell, selected structured inner
corrections in the same shell, and sparse sums of coordinate degree-seven
monomials.

\subsection{Exclusions for two restricted four-block families}
\label{subsec:four-block-residuals}

We first examine two restricted families arising from the four-block
construction in Lemma~\ref{lem:constg}.  The first is obtained by setting
\(g_3=0\).  The second restricts the two middle blocks to minimum-weight codewords 
whose sum has weight \(126\), while allowing the final correction block to vary 
in $\RM(5, 12)$.

Lou and Wang \cite{LouWang2026TwoConjectures} proved that
their Construction~1, based on specified sums of degree-six
monomials with constant block offsets, cannot produce a
codeword of weight \(322\) in \(\RM(7,14)\).
Their result concerns this particular construction and does
not establish the nonexistence of weight \(322\) in the full code.
Here, Proposition~\ref{prop:direct-family-exclusion} excludes
all four weights in \(T\) for the direct four-block family
with arbitrary \(g_1,g_2\in\RM(6,12)\), while
Proposition~\ref{prop:a64-126-exclusion} treats a separate
minimum-weight branch allowing a correction in \(\RM(5,12)\).
Both results remain restricted-family exclusions.

\begin{proposition}
\label{prop:direct-family-exclusion}
Let $ \mathcal F_{\rm dir} = \{0\|g_1\|g_2\|(g_1+g_2) \mid g_1,g_2\in\RM(6,12)\}.$
Then \(\mathcal F_{\rm dir}\) contains no codeword of weight in \(T\).
\end{proposition}

\begin{proof}
For \(G=0\|g_1\|g_2\|(g_1+g_2)\), one has $\wt(G)=\wt(g_1)+\wt(g_2)+\wt(g_1+g_2).$
If one of \(g_1,g_2,g_1+g_2\) is zero, then
\(\wt(G)=2\wt(h)\) for some \(h\in\RM(6,12)\).  A target weight would
therefore require $\wt(h)\in\{161,163,165,167\},$
which is impossible by the low-weight spectrum of \(\RM(6,12)\).

Otherwise \(g_1,g_2,g_1+g_2\) are the three nonzero elements of a
two-dimensional subspace of \(\RM(6,12)\).  The finite list of possible
triples and their exclusions are given in
Appendix~\ref{app:direct-four-block-inputs}, Proposition~\ref{prop:direct-triple-exclusion}.
Hence no codeword in \(\mathcal F_{\rm dir}\) has weight in \(T\).
\end{proof}

We next let the last block vary in the coset \(s_{126}+\RM(5,12)\), while
restricting the two middle blocks to minimum-weight codewords. Let \(A,B\subseteq\mathbb F_2^{12}\)
be affine \(6\)-flats with \(|A\cap B|=1\), and set
$ s_{126}=\mathbf 1_A+\mathbf 1_B, S=\supp(s_{126})=A\triangle B.$
Then \(\wt(s_{126})=126\).  All ordered pairs of affine \(6\)-flats meeting in
one point are affinely equivalent, so one fixed pair \(A,B\) suffices.  Define
$ \mathcal F_{\rm A64}^{(126)} = \{0\|\mathbf 1_A\|\mathbf 1_B\|(s_{126}+g_3)\mid g_3\in\RM(5,12)\},$
where the subscript records the two weight-\(64\) middle blocks and the
superscript records the weight of their sum.

For \(G\in\mathcal F_{\rm A64}^{(126)}\), the two middle blocks contribute \(128\).
Thus weights in \(T\) correspond to $\tau=\wt(s_{126}+g_3)\in\{194,198,202,206\}.$
Put $    \mu=|\supp(g_3)\cap S^c|, \rho=|\supp(g_3)\cap S|.$
Then $ \tau=126+\mu-\rho,\rho=\mu+126-\tau$ and $\wt(g_3)=2\mu+126-\tau .$

\begin{proposition}[The weight-\(126\) minimum-weight branch]
\label{prop:a64-126-exclusion}
The family \(\mathcal F_{\rm A64}^{(126)}\) contains no codeword of weight in \(T\).
By affine equivalence, this fixed-pair exclusion applies to every ordered
one-point-intersection pair of minimum-weight middle blocks.
\end{proposition}

\begin{proof}
By Appendix~\ref{app:a64-row-slice-summary},
Proposition~\ref{prop:a64-row-slice-reduction}, a target codeword in this family
can occur only when $\mu\in\{132,136,140\}.$
For \(\mu=140\), the possible values of \(\wt(g_3)\) are
\(212,208,204,200\), none of which occurs by
Lemma~\ref{lem:a64-rm512-low-input}.  For \(\mu=136\) and \(\mu=132\), the
same low-weight input leaves only $ (\tau,\wt(g_3),\rho)=(206,192,56)$ and $(198,192,60).$

By Lemma~\ref{lem:a64-rm512-low-input}, every weight-\(192\) codeword of
\(\RM(5,12)\) has the form \(\mathbf 1_C+\mathbf 1_D\), where \(C,D\) are
affine \(7\)-flats with \(\dim(C\cap D)=5\).  The exact finite overlap enumeration in 
Appendix~\ref{app:a64-weight192-verification},
Proposition~\ref{prop:a64-weight192-overlap}, gives
$|(C\triangle D)\cap S|\notin\{56,60\}$
for every relative position of such \(C,D\) with respect to the fixed pair
\(A,B\).  Thus the two remaining cases are also impossible for the fixed
normalized pair.  Since all ordered one-point-intersection pairs \(A,B\) are
affinely equivalent and affine transformations preserve Reed--Muller codes and
Hamming weight, the same exclusion holds for every one-point-intersection
minimum-weight branch.
\end{proof}

\begin{remark}
\label{rem:four-block-boundary}
Proposition~\ref{prop:direct-family-exclusion} excludes the direct
\(g_3=0\) four-block family, while
Proposition~\ref{prop:a64-126-exclusion} excludes the weight-\(126\)
one-point-intersection A64 branch.  These results show that the most immediate 
four-block extensions considered here do not produce the four target weights.
Thus the direct branch and the normalized weight-126 minimum-weight
branch do not yield the residual targets.  Other \(g_3\)-coset branches
require separate profile analyses.
\end{remark}

\subsection{Affine-flat corrections to a fixed Carlet-type Maiorana--McFarland outer shell}
\label{subsec:carlet-punctured}

Let \(V_6=\mathbb F_2^6\), \(V=V_6\times V_6\), and $Z=(V_6\setminus\{0\})\times(V_6\setminus\{0\}).$
Let $\delta_0(u)=\prod_{i=1}^6(u_i+1),$
so that \(\delta_0\) is supported at \(0\in V_6\).  Define the fixed
Carlet--Maiorana outer shell
\[
\Phi(x_1,x_2,u,v)
=
x_1\delta_0(u)+x_2\delta_0(v).
\]
For a correction \(g:V\to\mathbb F_2\), set $F_g(x_1,x_2,u,v)=\Phi(x_1,x_2,u,v)+g(u,v).$
If \(g\in\RM(7,12)\), then \(F_g\in\RM(7,14)\).  In particular, the
indicator of an affine \(5\)-flat in \(V\) is an admissible correction, as
is any finite sum of such indicators.

\begin{lemma}[Outer-shell reduction]
\label{lem:outer-shell-reduction}
For every correction \(g:V\to\mathbb F_2\),
\[
        \wt(F_g)=254+4\wt(g|_Z).
\]
\end{lemma}

\begin{proof}
For fixed \((u,v)\in V\), the fiber over \((x_1,x_2)\in\mathbb F_2^2\) is
\(x_1\delta_0(u)+x_2\delta_0(v)+g(u,v)\).  If \((u,v)\in Z\), this fiber is
constant and contributes \(4g(u,v)\).  Otherwise it is nonconstant and
contributes \(2\).  Since \(|V\setminus Z|=64+64-1=127\), the formula follows.
\end{proof}

Within the family \(F_g=\Phi+g\) generated from the fixed outer shell
\(\Phi\), Lemma~\ref{lem:outer-shell-reduction} gives
\(\wt(F_g)\in T\) if and only if
\(\wt(g|_Z)\in T_{\rm red}\), where
\(T_{\rm red}=\{17,18,19,20\}\).
Equivalently, $T=\{254+4t \mid t\in T_{\rm red}\}.$
Let $H_1=V_6\times\{0\},
H_2=\{0\}\times V_6,
H=H_1\cup H_2,$
so that \(Z=V\setminus H\).

\begin{lemma}[Punctured affine-flat values]
\label{lem:punctured-values}
If \(A\subseteq V\) is an affine \(5\)-flat, then
\[
        |A\cap Z|\in
        P_5:=\{0,15,16,21,22,23,24,25,26,27,28,29,30,31,32\}.
\]
If \(C\subseteq V\) is an affine \(3\)-flat, then
$|C\cap Z|\in P_3:=\{0,3,4,5,6,7,8\}.$
\end{lemma}

\begin{proof}
See Appendix~\ref{app:carlet-punctured-flat}.
\end{proof}

\begin{proposition}[Exclusion of affine \(5\)-flat correction families]
\label{prop:fixed-shell-affine-flat-exclusion}
For the family \(F_g=\Phi+g\) associated with the fixed outer shell
\(\Phi\), none of the following correction families produces a word of
weight in \(T\):
\(g=\mathbf 1_A\);
\(g=\sum_{i=1}^s\mathbf 1_{A_i}\), where \(s\ge2\) and the \(A_i\) are
pairwise disjoint; or
\(g=\mathbf 1_A+\mathbf 1_B\), where \(A\ne B\) and
\(A\cap B\ne\varnothing\).
Here \(A,B,A_i\subseteq V\) are affine \(5\)-flats.
\end{proposition}

\begin{proof}
By Lemma~\ref{lem:outer-shell-reduction}, it suffices to exclude reduced
weights in \(T_{\rm red}\).  For \(g=\mathbf 1_A\), the conclusion follows from
$P_5\cap T_{\rm red}=\varnothing.$

Now let $g=\sum_{i=1}^s\mathbf 1_{A_i}, s\ge2,$
where the \(A_i\) are pairwise disjoint.  Since their supports are disjoint,
there is no cancellation among the restrictions
\(\mathbf 1_{A_i}|_Z\).  If all these restrictions are zero, then
\(\wt(g|_Z)=0\).  If exactly one is nonzero, then
\(\wt(g|_Z)\) is a positive element of \(P_5\).  If at least two are
nonzero, then $\wt(g|_Z)\ge15+15=30.$
In every case, \(\wt(g|_Z)\notin T_{\rm red}\).

It remains to consider
\(g=\mathbf 1_A+\mathbf 1_B\), where
\(A\ne B\) and \(A\cap B\ne\varnothing\).
Put \(C=A\cap B\).  Since \(A\) and \(B\) are distinct affine
\(5\)-flats with nonempty intersection, one has
$ \dim C\in\{0,1,2,3,4\}. $
The cases \(\dim C\in\{0,1,2,4\}\) are excluded by
Lemma~\ref{lem:carlet-coarse-intersection-cases}.

It remains to consider \(\dim C=3\).  Put
$a=|A\cap Z|, b=|B\cap Z|, c=|C\cap Z|.$
Then $\wt((\mathbf 1_A+\mathbf 1_B)|_Z)=a+b-2c$, where $a,b\in P_5, c\in P_3.$
Lemma~\ref{lem:carlet-d3-candidate-screening} reduces the possible triples
\((a,b,c)\) to Table~\ref{tab:carlet-d3-candidates}, and
Lemma~\ref{lem:carlet-d3-local-exclusion} excludes all triples in that table.
Hence no reduced weight in \(T_{\rm red}\) occurs.
\end{proof}

\begin{remark}[Product-type subfamily]
\label{rem:product-type-pairs}
Product-type affine \(5\)-flats \(A=X\times Y\), with
\(\dim X+\dim Y=5\), form a separable subfamily of the affine
\(5\)-flat corrections already covered by
Proposition~\ref{prop:fixed-shell-affine-flat-exclusion}.  A direct
block-preserving product calculation gives the same no-hit conclusion for
\(T_{\rm red}=\{17,18,19,20\}\); this serves only as a consistency check and
is not used in the proof.
\end{remark}

\begin{remark}
\label{rem:fixed-shell-boundary}
Proposition~\ref{prop:fixed-shell-affine-flat-exclusion} excludes,
within the fixed Carlet-type Maiorana--McFarland outer shell \(\Phi\),
the following explicitly stated correction families:
a single affine \(5\)-flat;
sums of at least two pairwise-disjoint affine \(5\)-flats; and
sums of two distinct affine \(5\)-flats with nonempty intersection.
All of these families avoid the reduced target set $T_{\rm red}=\{17,18,19,20\}. $
The product-type pair calculation in
Remark~\ref{rem:product-type-pairs} is contained in these cases and
serves only as a consistency check.  No claim is made here for sums of
three or more affine \(5\)-flats having nontrivial mutual
intersections, or for affine-flat corrections outside the fixed outer
shell \(\Phi\).
\end{remark}

\subsection{Structured inner corrections in the fixed Carlet-type
Maiorana--McFarland outer shell}
\label{subsec:carlet-inner-mm}

We continue with the family \(F_g=\Phi+g\) associated with the fixed
Carlet-type Maiorana--McFarland outer shell \(\Phi\) from
Subsection~\ref{subsec:carlet-punctured}.  Put
$ U=V_6\setminus\{0\}, $ and $Z=U\times U.$

By Lemma~\ref{lem:outer-shell-reduction},
$ \wt(F_g)=254+4\wt(g|_Z), $
so the target set \(T=\{322,326,330,334\}\) corresponds to
$ T_{\rm red}=\{17,18,19,20\}.$
Appendix~\ref{app:inner-mm-refinements} gives a self-contained row-degree
analysis and intrinsic definitions of the pure-cross, low-block-rank, and
mixed-core subfamilies used below.

For \(q:U\to\mathbb F_2\), write
$ \wt_U(q)=|\{v\in U \mid q(v)=1\}|, \delta(q)=\min\{\wt_U(q),63-\wt_U(q)\}.$

\begin{lemma}[Row-wise elimination]
\label{lem:row-wise-elimination}
Let \(P:U\times U\to\mathbb F_2\) be fixed.  Then
\[
\min_{h:U\to\mathbb F_2}
\wt\bigl((u,v)\mapsto h(u)+P(u,v)\bigr)
=
\sum_{u\in U}\delta(P(u,\cdot)).
\]
\end{lemma}

\begin{proof}
For each fixed \(u\), the value \(h(u)\) either keeps the row
\(P(u,\cdot)\) or complements it on the \(63\)-point set \(U\).  The smaller
of the two row contributions is \(\delta(P(u,\cdot))\), and the choices are
independent from row to row.
\end{proof}

The next proposition is stronger than the individual low-complexity searches:
it closes the entire correction space attached to this one fixed outer shell.
It remains a restricted-family result in \(\RM(7,14)\), because the outer
shell \(\Phi\) is fixed.

\begin{proposition}[Complete exclusion in the fixed outer shell]
\label{prop:fixed-shell-complete-exclusion}
For every \(g\in\RM(7,12)\), $ \wt(F_g)\notin T.$
Equivalently, $ \wt(g|_Z)\notin T_{\rm red}.$
\end{proposition}

\begin{proof}
Group the algebraic normal form of \(g(u,v)\) by its \(v\)-monomials.  On the
punctured set \(U\), a degree-six \(v\)-monomial can be replaced by a
polynomial of degree at most five using
\[
\prod_{j=1}^{6}(1+v_j)=0.
\]
Thus \(g|_Z\) has a degree-compatible separated presentation of total degree
at most \(7\) in the sense of
\eqref{eq:separated-presentation}.  Proposition~\ref{prop:app-separated}
excludes all four values in \(T_{\rm red}\), and
Lemma~\ref{lem:outer-shell-reduction} then gives the asserted exclusion for
\(F_g\).
\end{proof}

\begin{corollary}[Pure-cross and low-block-rank refinements]
\label{prop:pure-cross-separated-exclusion}
The following explicitly defined correction families avoid
\(T_{\rm red}\):
\begin{enumerate}
\item nonzero sums of one to five allowed degree-seven cross factors from
\eqref{eq:allowed-cross-factor};
\item the reduced block-rank families \(\mathscr L_2\) and \(\mathscr L_3\)
defined in \eqref{eq:L2-L3-definition};
\item the normalized weight-\(15\) and weight-\(16\) base-perturbation
families \(\mathscr F_{15}^{(\le3)}\) and \(\mathscr F_{16}^{(<5)}\) defined
in \eqref{eq:F15-family}--\eqref{eq:F16-family}.
\end{enumerate}
Moreover, in the \(v\)-distinct pure-cross profile subfamily, the relaxed
row minimum is at least \(31\), and this bound is attained.
\end{corollary}

\begin{proof}
The target exclusions follow from
Proposition~\ref{prop:fixed-shell-complete-exclusion}, or directly from
Proposition~\ref{prop:app-separated}.  The quantitative pure-cross statement
is Proposition~\ref{prop:app-pure-cross}.
\end{proof}

We retain the mixed-core terminology because it records useful affine
geometry not visible in the row-degree proof.  Let
\(S=a+L\subseteq V_6\times V_6\) be an affine \(5\)-flat and put
\(X=S\cap Z\).  With $ H_1=V_6\times\{0\}$ and $H_2=\{0\}\times V_6,$
define $r_1=\dim(L\cap H_1),r_2=\dim(L\cap H_2)
$ and $ k=5-r_1-r_2.$
We call \(X\) genuinely mixed when \(k>0\).

\begin{corollary}[Single-, two-, and three-core mixed corrections]
\label{prop:mixed-affine-core-exclusion}
Within the fixed outer shell \(\Phi\), none of the following corrections
produces a reduced weight in \(T_{\rm red}\):
\begin{enumerate}
\item a single genuinely mixed affine \(5\)-flat core;
\item the sum of two genuinely mixed affine \(5\)-flat cores;
\item the sum of three genuinely mixed affine \(5\)-flat cores, namely every
member of the family \(\mathscr M_3\) defined in \eqref{eq:M3-definition}.
\end{enumerate}
\end{corollary}

\begin{proof}
The single-core reduced sizes are given in
Lemma~\ref{lem:single-core-values}.  The two-core case is also covered by
Proposition~\ref{prop:fixed-shell-affine-flat-exclusion}.  The complete
three-core statement is
Proposition~\ref{prop:three-core-residual-exclusions}.  Alternatively, all
three cases follow at once from
Proposition~\ref{prop:fixed-shell-complete-exclusion}.
\end{proof}

\begin{remark}
\label{rem:inner-correction-boundary}
Proposition~\ref{prop:fixed-shell-complete-exclusion} is complete with respect
to the inner correction \(g\in\RM(7,12)\) while the Carlet-type outer shell
\(\Phi\) is fixed.  It does not assert that the four weights in \(T\) are
absent from the full code \(\RM(7,14)\), because a general codeword of
\(\RM(7,14)\) need not belong to this fixed-shell family.  The pure-cross,
block-rank, base-perturbation, and affine-core statements above are retained
as quantitative or geometric refinements of this fixed-shell exclusion.
\end{remark}

\subsection{Sparse sums of degree-\(7\) coordinate monomials}
\label{subsec:sparse-degree-seven-monomial-sums}

Recall that \([14]=\{1,\ldots,14\}\).  For
\(I\in\binom{[14]}{7}\), write $ m_I=\prod_{i\in I}x_i\in B_{14}.$
We consider sums of squarefree coordinate monomials of degree \(7\).  Since the
sum is over \(\mathbb F_2\), repeated monomials cancel in pairs, so it is
enough to consider sums of distinct monomials.  For \(s\ge0\), define
\[
        \mathcal W_s^{\rm dist}
        =
        \left\{
        \wt\left(\sum_{I\in\mathcal I}m_I\right)\mid
        \mathcal I\subseteq\binom{[14]}{7},\ |\mathcal I|=s
        \right\}.
\]
The case \(s=0\) gives the zero function.

\begin{proposition}[Exclusion for sums of at most eight degree-\(7\) coordinate monomials]
\label{prop:sparse-coordinate-monomial-exclusion}
Let \(T=\{322,326,330,334\}\).  Then, for every \(0\le s\le8\),
$ \mathcal W_s^{\rm dist}\cap T=\varnothing. $
Consequently, no sum of at most eight squarefree coordinate monomials of
degree \(7\), allowing repetitions before cancellation, has weight in \(T\).
\end{proposition}

\begin{proof}
The support of \(m_I\) is the affine \(7\)-flat $ A_I=\{x\in\mathbb F_2^{14} \mid x_i=1\text{ for all }i\in I\}. $
For distinct \(I_1,\ldots,I_s\), the support of
\(\sum_{r=1}^s m_{I_r}\) is the symmetric difference of
\(A_{I_1},\ldots,A_{I_s}\).  Hence
\[
\wt\left(\sum_{r=1}^s m_{I_r}\right)
=
\sum_{\emptyset\ne J\subseteq[s]}
(-2)^{|J|-1}
2^{14-\left|\bigcup_{j\in J}I_j\right|}.
\]
Proposition~\ref{prop:sparse7-low-term} excludes target weights for
\(0\le s\le4\), and Proposition~\ref{prop:sparse7-s5s7-enumeration} excludes
them for \(5\le s\le7\).

It remains to consider \(s=8\).  By Lemma~\ref{lem:sparse7-fixed-pair}, any
eight-term target configuration must contain a complementary pair of
\(7\)-subsets.  After normalizing such a pair to
$ \{1,\ldots,7\}, \{8,\ldots,14\},$
Proposition~\ref{prop:sparse7-fixed-pair-enumeration} excludes all remaining
fixed-pair profiles.  Thus \(\mathcal W_s^{\rm dist}\cap T=\varnothing\) for
every \(0\le s\le8\).

Finally, any expression with repetitions and \(t\le8\) reduces over
\(\mathbb F_2\) to a sum of \(s\le8\) distinct monomials.  The distinct-summand
case therefore proves the claim.
\end{proof}

\begin{remark}
\label{rem:sparse-coordinate-boundary}
Proposition~\ref{prop:sparse-coordinate-monomial-exclusion} concerns only
coordinate squarefree degree-\(7\) monomials.  It does not apply to general
affine \(7\)-flat indicators, arbitrary affine changes of variables,
lower-degree corrections, or general highest-degree cosets
\(h_7+\RM(6,14)\). Thus it should be read as an exclusion for sparse sums of
coordinate degree-\(7\) monomials, not as an exclusion for sums of arbitrary
minimum-weight affine-\(7\)-flat indicators.
\end{remark}

\begin{remark}[Other structured routes]
We also considered three further structured routes for the residual weights.
The first is a highest-degree skeleton analysis of the cosets
\(h_7+\RM(6,14)\).  The second is the affine \(7\)-flat assembly problem
beyond the coordinate-monomial subfamily considered above.  The third is a
global Maiorana--McFarland core-pooling approach.  Since these broader
families have not been exhausted, we do not derive additional exclusion
results from them here and leave their complete treatment for future work.
\end{remark}

\section{Conclusion}
\label{sec:conclusion}

We established a constructive inclusion for the weight spectrum of the
high-order Reed--Muller family \(\RM(m-7,m)\).  For every \(m\ge 14\),
the spectrum \(W(m-7,m)\) contains all even weights in
\([336,2^m-336]\), together with the finite low-weight set \(A\) in
Theorem~\ref{thm:main-thm} and the complements of those weights.  The proof
is based on an explicit base-code construction in \(\RM(7,14)\), using
four-block concatenation, degree-six monomial witnesses, and known
spectrum intervals for \(\RM(5,10)\) and \(\RM(6,12)\).  A
concatenation-based induction then lifts the base inclusion to all
\(m\ge 14\).

We also investigated the four remaining low base-code target weights
\(T=\{322,326,330,334\}\) through several restricted construction
families. The resulting exclusions cover two four-block branches,
all corrections \(g\in\RM(7,12)\) to a fixed Carlet-type
Maiorana--McFarland outer shell, and sums of at most eight
squarefree coordinate monomials of degree \(7\).
These results provide further structural information on the
remaining spectrum problem. Determining whether the weights in
\(T\) occur in the full code \(\RM(7,14)\) remains open.

\appendix
\section{Exact verification and witness table for the base interval}
\label{app:base-interval-witnesses}
This appendix provides a representative exact weight calculation for one of
the monomial constructions used in Section~\ref{sec:key-section}.  It also
records the remaining monomial witnesses used in
Proposition~\ref{prop:base_interval}.

For \(I\in\binom{[12]}{6}\), write
\(m_I=\prod_{i\in I}x_i\). All sums are taken over \(\mathbb F_2\).
For a sum \(f=\sum_{j=1}^{r}m_{I_j}\), define
\[
S_k(f)
=
\sum_{\substack{J\subseteq\{1,\ldots,r\}\\ |J|=k}}
2^{12-\left|\bigcup_{j\in J}I_j\right|}.
\]
Since the support of \(f\) is the symmetric difference of the supports of
its constituent monomials, inclusion--exclusion for symmetric differences
gives $\wt(f)=\sum_{k=1}^{r}(-2)^{k-1}S_k(f).$

\subsection{A representative exact calculation}
\label{app:354}

For the weight-\(354\) construction, let
\[
\begin{aligned}
m_1&=x_3x_4x_7x_8x_{11}x_{12},&
m_2&=x_3x_4x_6x_7x_8x_{11},\\
m_3&=x_1x_3x_4x_7x_8x_{11},&
m_4&=x_1x_4x_5x_8x_9x_{11},\\
m_5&=x_2x_5x_6x_9x_{10}x_{12},&
m_6&=x_2x_3x_5x_6x_9x_{12}.
\end{aligned}
\]
Set $g_1=m_1+m_2+m_3,$ and $ g_2=m_4+m_5+m_6.$
A direct calculation gives
\[
\begin{aligned}
\bigl(S_1(g_1),S_2(g_1),S_3(g_1)\bigr)
&=(192,96,16),\\
\bigl(S_1(g_2),S_2(g_2),S_3(g_2)\bigr)
&=(192,40,2),\\
\bigl(S_1(g_1+g_2),\ldots,S_6(g_1+g_2)\bigr)
&=(384,183,64,23,7,1).
\end{aligned}
\]
Consequently,
\[
\begin{aligned}
\wt(g_1)
&=192-2\cdot96+4\cdot16=64,\\
\wt(g_2)
&=192-2\cdot40+4\cdot2=120,\\
\wt(g_1+g_2)
&=384-2\cdot183+4\cdot64-8\cdot23
  +16\cdot7-32\cdot1=170.
\end{aligned}
\]
Therefore, $\wt\bigl(0\,\|\,g_1\,\|\,g_2\,\|\,(g_1+g_2)\bigr)
=64+120+170=354.$

\subsection{Remaining monomial witnesses}
\label{app:remaining-monomial-witnesses}

To save space in Table~\ref{tab:base-interval-witnesses}, we write
\(m_{1,2,3,4,5,6}\) for \(m_{\{1,2,3,4,5,6\}}\), and similarly for
the other monomials.  For every row, the word
\(0\|g_1\|g_2\|(g_1+g_2)\) has the target weight.

\begingroup
\tiny
\setlength{\tabcolsep}{2pt}
\renewcommand{\arraystretch}{1.13}
\setlength{\LTleft}{0pt}
\setlength{\LTright}{0pt}

\begin{longtable}{@{}c|p{0.335\textwidth}|p{0.335\textwidth}|c@{}}
\caption{Remaining monomial witnesses for Proposition~\ref{prop:base_interval}.  Together with the five witnesses
displayed in the proof of Proposition~\ref{prop:base_interval}, these rows cover all target weights congruent to \(2\pmod4\) in
\([338,474]\cup[538,542]\).}
\label{tab:base-interval-witnesses}\\
\toprule
Target & \(g_1\) & \(g_2\) &
\((\wt(g_1),\wt(g_2),\wt(g_1+g_2))\)\\
\midrule

\endfirsthead

\multicolumn{4}{@{}l}{%
  \textit{\tablename~\thetable\ (continued)}}\\[2pt]
\toprule
Target & \(g_1\) & \(g_2\) &
\((\wt(g_1),\wt(g_2),\wt(g_1+g_2))\)\\
\midrule
\endhead

\midrule
\multicolumn{4}{r}{Continued on next page}\\
\endfoot

\bottomrule
\endlastfoot

\(342\)
& \(\biwthree{1,2,3,4,5,6}{1,2,3,4,5,7}{1,2,3,4,8,9}\)
& \(\biwthree{1,2,3,4,5,6}{1,2,3,4,5,7}{6,8,9,10,11,12}\)
& \((96,126,120)\)
\\

\(346\)
& \(\biwthree{1,2,3,4,5,6}{1,2,3,4,5,7}{1,2,3,4,7,8}\)
& \(\biwthree{1,2,3,4,5,6}{1,2,3,4,5,7}{5,6,9,10,11,12}\)
& \((96,124,126)\)
\\

\(350\)
& \(\biwthree{1,2,3,4,5,6}{1,2,3,4,5,7}{1,2,3,4,8,9}\)
& \(\biwthree{1,2,3,4,5,6}{1,2,3,4,5,7}{5,6,7,10,11,12}\)
& \((96,128,126)\)
\\

\(358\)
& \(\biwthree{1,2,3,4,5,6}{1,2,3,4,5,7}{1,2,3,7,8,9}\)
& \(\biwthree{1,2,3,4,5,6}{1,2,3,4,5,7}{4,5,6,10,11,12}\)
& \((112,120,126)\)
\\

\(362\)
& \(\biwthree{1,2,3,4,5,6}{3,4,5,9,10,11}{3,4,5,6,9,11}\)
& \(\biwthree{1,2,3,4,5,6}{7,8,9,10,11,12}{2,7,8,9,10,12}\)
& \((112,126,124)\)
\\

\(366\)
& \(\biwthree{1,2,3,4,5,6}{1,2,3,4,5,7}{1,2,3,8,9,10}\)
& \(\biwthree{1,2,3,4,5,6}{1,2,3,4,5,7}{4,5,6,7,11,12}\)
& \((112,128,126)\)
\\

\(370\)
& \(\biwthree{1,2,3,4,5,6}{1,2,3,4,7,8}{1,2,3,4,7,9}\)
& \(\biwthree{1,2,3,4,5,6}{1,2,3,4,7,8}{7,8,9,10,11,12}\)
& \((96,154,120)\)
\\

\(374\)
& \(\biwthree{1,2,3,4,5,6}{1,2,3,4,7,8}{1,2,3,4,7,9}\)
& \(\biwthree{1,2,3,4,5,6}{1,2,3,4,7,8}{5,6,9,10,11,12}\)
& \((96,154,124)\)
\\

\(378\)
& \(\biwthree{1,2,3,4,5,6}{1,2,3,4,7,8}{1,2,3,4,7,9}\)
& \(\biwthree{1,2,3,4,5,6}{1,2,3,4,7,8}{5,6,8,10,11,12}\)
& \((96,156,126)\)
\\

\(382\)
& \(\biwthree{1,2,3,4,5,6}{1,2,5,8,9,11}{2,5,6,8,9,11}\)
& \(\biwthree{1,2,3,4,5,6}{3,4,6,7,10,12}{1,3,6,7,10,12}\)
& \((128,128,126)\)
\\

\(386\)
& \(\biwthree{1,2,3,4,5,6}{1,2,3,7,8,9}{1,2,3,7,8,10}\)
& \(\biwthree{1,2,3,4,5,6}{1,2,3,7,8,9}{7,8,9,10,11,12}\)
& \((112,162,112)\)
\\

\(390\)
& \(\biwthree{1,2,3,4,5,6}{1,2,3,4,7,8}{1,2,3,7,8,9}\)
& \(\biwthree{1,2,3,4,5,6}{1,2,3,4,7,8}{5,6,9,10,11,12}\)
& \((112,154,124)\)
\\

\(394\)
& \(\biwthree{1,2,3,4,5,6}{1,2,3,7,8,9}{1,2,3,4,7,8}\)
& \(\biwthree{1,2,3,4,5,6}{1,2,3,7,8,9}{7,8,9,10,11,12}\)
& \((112,162,120)\)
\\

\(398\)
& \(\biwthree{1,2,3,4,5,6}{1,2,3,7,8,9}{1,2,3,7,8,10}\)
& \(\biwthree{1,2,3,4,5,6}{1,2,3,7,8,9}{4,5,6,10,11,12}\)
& \((112,162,124)\)
\\

\(402\)
& \(\biwthree{1,2,3,4,5,6}{1,2,3,7,8,9}{1,2,3,7,8,10}\)
& \(\biwthree{1,2,3,4,5,6}{1,2,3,7,8,9}{4,5,6,9,11,12}\)
& \((112,164,126)\)
\\

\(406\)
& \(\biwthree{1,2,3,4,5,6}{1,2,3,7,8,9}{1,2,3,4,7,8}\)
& \(\biwthree{1,2,3,4,5,6}{1,2,3,7,8,9}{5,6,9,10,11,12}\)
& \((112,168,126)\)
\\

\(410\)
& \(\biwthree{1,2,3,4,5,6}{1,2,3,7,8,9}{1,2,3,4,7,10}\)
& \(\biwthree{1,2,3,4,5,6}{1,2,3,7,8,9}{4,5,6,10,11,12}\)
& \((128,162,120)\)
\\

\(414\)
& \(\biwthree{1,2,3,4,5,6}{1,2,7,8,9,10}{1,2,3,7,8,9}\)
& \(\biwthree{1,2,3,4,5,6}{1,2,7,8,9,10}{4,5,6,10,11,12}\)
& \((120,168,126)\)
\\

\(418\)
& \(\biwthree{1,2,3,4,5,6}{1,2,3,7,8,9}{1,2,3,7,10,11}\)
& \(\biwthree{1,2,3,4,5,6}{1,2,3,7,8,9}{4,5,6,10,11,12}\)
& \((136,162,120)\)
\\

\(422\)
& \(\biwthree{1,2,3,4,5,6}{1,2,3,7,8,9}{1,2,3,4,7,10}\)
& \(\biwthree{1,2,3,4,5,6}{1,2,3,7,8,9}{5,6,8,9,11,12}\)
& \((128,168,126)\)
\\

\(426\)
& \(\biwthree{1,2,3,4,5,6}{1,2,3,7,8,9}{1,2,7,8,10,11}\)
& \(\biwthree{1,2,3,4,5,6}{1,2,3,7,8,9}{4,5,6,10,11,12}\)
& \((144,162,120)\)
\\

\(430\)
& \(\biwthree{1,2,3,4,5,6}{1,2,3,7,8,9}{1,2,3,7,10,11}\)
& \(\biwthree{1,2,3,4,5,6}{1,2,3,7,8,9}{4,5,6,8,9,12}\)
& \((136,168,126)\)
\\

\(434\)
& \(\biwthree{1,2,3,4,5,6}{1,2,7,8,9,10}{1,2,3,4,7,8}\)
& \(\biwthree{1,2,3,4,5,6}{1,2,7,8,9,10}{5,6,9,10,11,12}\)
& \((136,172,126)\)
\\

\(438\)
& \(\biwthree{1,2,3,4,5,6}{1,2,7,8,9,10}{1,2,3,7,8,11}\)
& \(\biwthree{1,2,3,4,5,6}{1,2,7,8,9,10}{4,5,6,9,10,12}\)
& \((144,168,126)\)
\\

\(442\)
& \(\biwthree{1,2,3,4,5,6}{1,7,8,9,10,11}{1,2,3,7,8,9}\)
& \(\biwthree{1,2,3,4,5,6}{1,7,8,9,10,11}{4,5,6,10,11,12}\)
& \((148,168,126)\)
\\

\(446\)
& \(\biwthree{1,2,3,4,5,6}{1,2,7,8,9,10}{1,3,4,7,8,11}\)
& \(\biwthree{1,2,3,4,5,6}{1,2,7,8,9,10}{2,5,6,9,10,12}\)
& \((160,160,126)\)
\\

\(450\)
& \(\biwthree{1,2,3,4,5,6}{1,2,3,7,8,9}{1,2,3,10,11,12}\)
& \(\biwthree{1,2,3,4,5,6}{1,2,3,7,8,9}{4,5,6,7,8,9}\)
& \((148,176,126)\)
\\

\(454\)
& \(\biwthree{1,2,3,4,5,6}{1,4,6,8,9,11}{2,5,6,9,10,12}\)
& \(\biwthree{1,2,3,4,5,6}{1,3,4,7,8,11}{1,3,4,6,7,11}\)
& \((160,112,182)\)
\\

\(458\)
& \(\biwthree{1,2,3,4,5,6}{1,2,5,6,7,10}{2,3,5,8,9,10}\)
& \(\biwthree{1,2,3,4,5,6}{2,3,5,6,7,10}{3,4,8,9,11,12}\)
& \((144,152,162)\)
\\

\(462\)
& \(\biwthree{1,2,3,4,5,6}{1,3,4,5,9,11}{1,3,4,9,10,12}\)
& \(\biwthree{1,2,3,4,5,6}{2,3,4,6,9,11}{7,8,9,10,11,12}\)
& \((128,154,180)\)
\\

\(466\)
& \(\biwthree{1,2,3,4,5,6}{4,5,7,8,10,11}{4,6,7,8,10,12}\)
& \(\biwthree{1,2,3,4,5,6}{1,2,7,8,9,12}{7,8,9,10,11,12}\)
& \((152,154,160)\)
\\

\(470\)
& \(\biwthree{1,2,3,4,5,6}{4,5,7,8,10,12}{1,2,4,5,10,12}\)
& \(\biwthree{1,2,3,4,5,6}{7,8,9,10,11,12}{2,3,5,6,7,10}\)
& \((136,154,180)\)
\\

\end{longtable}
\endgroup

\section{Detailed exclusions for restricted four-block families}
\label{app:four-block-residuals}

Throughout this appendix let \(V=\mathbb F_2^{12}\), and let
\(T=\{322,326,330,334\}\).  For \(X\subseteq V\), write \(\mathbf 1_X\) for
its indicator function, and call a \(d\)-dimensional affine subspace of \(V\)
a \(d\)-flat.

\subsection{The direct four-block family}
\label{app:direct-four-block-inputs}

This subsection proves the finite-geometric exclusion used in
Proposition~\ref{prop:direct-family-exclusion}.  We consider the three
nonzero words of a two-dimensional subspace of \(\RM(6,12)\).

\begin{lemma}[Low-weight input for \(\RM(6,12)\)]
\label{lem:tool3-low-weight-normal-forms}
The weights of \(\RM(6,12)\) in the range \(0\le w\le 2048\) are
\[
\begin{aligned}
W(6,12)\cap[0,2048]
=\{0,64,96,112,120,124,126,128,136,144,148\}\cup \{152,154,156,\ldots,2048\}.
\end{aligned}
\]
Moreover, the following low-weight normal forms are used below.
\begin{enumerate}
\item If \(\wt(f)=64\), then \(f=\mathbf 1_M\) for an affine \(6\)-flat
\(M\subseteq V\).
\item If \(\wt(f)=96\), then \(f=\mathbf 1_A+\mathbf 1_B\), where \(A,B\)
are affine \(6\)-flats and \(A\cap B\) is an affine \(4\)-flat.
\item If \(\wt(f)=112\), then \(f\) is affinely equivalent to one of
$x_1x_2x_3(x_4x_5x_6+x_7x_8x_9)$ or $x_1x_2x_3x_4(x_5x_6+x_7x_8+x_9x_{10}).$
Here \(x_i\) denote affine coordinate functions after an affine change of
variables. The first type is the sum of two minimum-weight words whose supports 
meet in an affine \(3\)-flat; the second is the quadratic type.
\item If \(\wt(f)=126\), then \(f=\mathbf 1_A+\mathbf 1_B\), where \(A,B\)
are affine \(6\)-flats and \(A\cap B\) is one affine point.
\end{enumerate}
\end{lemma}

\begin{proof}
The spectrum statement follows from the known spectrum of \(\RM(m-6,m)\) at
\(m=12\).  The normal forms are the corresponding Kasami--Tokura low-weight
normal forms.  In the two-flat cases,
$ \wt(\mathbf 1_A+\mathbf 1_B) = |A\triangle B| = 128-2|A\cap B|.$
\end{proof}

\begin{lemma}[\(64\)-flat jump bound]
\label{lem:tool3-jump-bound}
Let \(M\subseteq V\) be an affine \(6\)-flat and let \(h\in\RM(6,12)\).  Then
$\wt(\mathbf 1_M+h)=64+\wt(h)-2|\supp(h)\cap M|.$
In particular, \(|\wt(\mathbf 1_M+h)-\wt(h)|\le 64\).
\end{lemma}

\begin{proof}
The support of \(\mathbf 1_M+h\) is \(M\triangle\supp(h)\).  The formula is
therefore immediate, and the inequality follows from
\(0\le |\supp(h)\cap M|\le64\).
\end{proof}

\begin{lemma}[Intersections of a \(6\)-flat with low-weight supports]
\label{lem:tool3-affine-intersection-obstruction}
Let \(M\subseteq V\) be an affine \(6\)-flat.
\begin{enumerate}
\item If \(h\in\RM(6,12)\) has weight \(126\), then
$|M\cap\supp(h)|\notin\{23,27\}.$
\item If \(h\in\RM(6,12)\) has weight \(112\), then
$|M\cap\supp(h)|\notin\{9,11\}.$
\end{enumerate}
\end{lemma}

\begin{proof}
For \(\wt(h)=126\), write \(h=\mathbf 1_A+\mathbf 1_B\), where
\(A,B\) are affine \(6\)-flats meeting in one point.  Then
$|M\cap\supp(h)|=|M\cap A|+|M\cap B|-2|M\cap A\cap B|.$
All nonzero terms are powers of two, so this expression cannot be \(23\) or
\(27\).

Let \(\wt(h)=112\).  If \(h\) is of the two-flat type, write
\(h=\mathbf 1_A+\mathbf 1_B\), where \(A\cap B\) is an affine \(3\)-flat.
The same power-of-two check excludes \(11\).  For the value \(9\), the only
possible numerical configuration, up to exchanging \(A\) and \(B\), is
$|M\cap A|=1, |M\cap B|=8, M\cap(A\cap B)=\varnothing.$

Translate by a point of \(A\cap B\), so that \(A,B\) and \(P=A\cap B\) are
linear, and write \(M=a+L\).  The equalities above give
\(\dim(L\cap A)=0\) and \(\dim(L\cap B)=3\).  Hence \(L\cap P=0\),
\(\dim(P+L)=9\), and \(\dim(B+L)=9\).  Since \(P+L\subseteq B+L\), the two
spaces are equal.  But \(M\cap B\ne\varnothing\), so \(a\in B+L=P+L\), which
forces \(M\cap P\ne\varnothing\), a contradiction.

It remains to consider the quadratic type.  Up to affine equivalence,
$h=x_1x_2x_3x_4(x_5x_6+x_7x_8+x_9x_{10}).$
The support of \(h\) lies in the affine \(8\)-flat
\(H=\{x_1=x_2=x_3=x_4=1\}\).  The intersection \(M\cap H\) is either empty or
an affine \(d\)-flat with \(d\le6\).  On this affine \(d\)-flat, the remaining
condition is a quadratic Boolean function.  By the standard weight formula for
quadratic Boolean functions, the possible numbers of ones are
\[
\begin{array}{c|c}
d & \text{possible numbers of ones}\\
\hline
0 & \{0,1\}\\
1 & \{0,1,2\}\\
2 & \{0,1,2,3,4\}\\
3 & \{0,2,4,6,8\}\\
4 & \{0,4,6,8,10,12,16\}\\
5 & \{0,8,12,16,20,24,32\}\\
6 & \{0,16,24,28,32,36,40,48,64\}.
\end{array}
\]
Neither \(9\) nor \(11\) occurs.
\end{proof}

\begin{lemma}[Intersections with a weight-\(96\) support]
\label{lem:tool3-two-triangle-values}
Let \(E=\supp(u)\), where \(u\in\RM(6,12)\) has weight \(96\).  Then, for
every affine \(6\)-flat \(X\subseteq V\),
\[
        |X\cap E|\in
        S_6:=\{0,1,2,3,4,6,8,12,16,24,32,48\}.
\]
\end{lemma}

\begin{proof}
Write \(u=\mathbf 1_A+\mathbf 1_B\), where \(P=A\cap B\) is an affine
\(4\)-flat.  After translating a point of \(P\) to the origin, \(P,A,B\) are
linear.  Let \(\pi:V\to V/P\).  In the \(4\)-space
\(W=(A+B)/P\subseteq V/P\), the image \(\pi(E)\) is the union of two disjoint
punctured \(2\)-spaces.

Let \(X=a+L\) be an affine \(6\)-flat and put \(t=\dim(L\cap P)\).  Then
\(\pi(X)\) has dimension \(6-t\).  If \(Z=\pi(X)\cap W\) is nonempty, it is an
affine \(s\)-flat with $\max\{0,2-t\}\le s\le \min\{4,6-t\}.$
A direct check in \(W\cong\mathbb F_2^4\) gives
\[
\begin{array}{c|c}
s & |Z\cap\pi(E)|\\
\hline
0 & \{0,1\}\\
1 & \{0,1,2\}\\
2 & \{0,1,2,3\}\\
3 & \{2,4\}\\
4 & \{6\}.
\end{array}
\]
Each point of \(Z\cap\pi(E)\) lifts to \(2^t\) points of \(X\cap E\).  Thus
\[
\begin{array}{c|c}
t & |X\cap E|\\
\hline
0 & \{0,1,2,3,4,6\}\\
1 & \{0,2,4,6,8,12\}\\
2 & \{0,4,8,12,16,24\}\\
3 & \{0,8,16,24,32\}\\
4 & \{0,16,32,48\}.
\end{array}
\]
Taking the union over \(t\) gives the stated set \(S_6\).
\end{proof}

\begin{proposition}[Direct four-block triple exclusion]
\label{prop:direct-triple-exclusion}
Let \(U\) be a two-dimensional subspace of \(\RM(6,12)\).  Then
$\sum_{u\in U\setminus\{0\}}\wt(u)\notin T.$
\end{proposition}

\begin{proof}
Arrange the three nonzero weights as \(w_1\le w_2\le w_3\).  Since every
nonzero word has weight at least \(64\), any triple with sum in \(T\) has all
entries at most \(206\).  The spectrum in
Lemma~\ref{lem:tool3-low-weight-normal-forms} gives the following candidates:
\[
\begin{array}{c|l}
322 & (64,64,194),\ (64,96,162)\\
326 & (64,64,198),\ (64,96,166),\ (64,126,136)\\
330 & (64,64,202),\ (64,96,170),\ (64,112,154)\\
334 & (64,64,206),\ (64,96,174),\ (64,112,158),\ (64,126,144),\
      (96,112,126).
\end{array}
\]
The triples \((64,64,w)\) and \((64,96,w)\) listed above violate
Lemma~\ref{lem:tool3-jump-bound}.  The triples \((64,126,136)\) and
\((64,126,144)\) require intersections \(27\) and \(23\) with a weight-\(126\)
support, and the triples \((64,112,154)\) and \((64,112,158)\) require
intersections \(11\) and \(9\) with a weight-\(112\) support.  These are
excluded by Lemma~\ref{lem:tool3-affine-intersection-obstruction}.

It remains to exclude \((96,112,126)\).  Choose the weight-\(96\) word and
the weight-\(126\) word as two nonzero elements of \(U\); their sum is then
the remaining weight-\(112\) word.  Let \(E\) be the support of the
weight-\(96\) word and let \(F=C\triangle D\) be the support of the
weight-\(126\) word, with \(C,D\) affine \(6\)-flats meeting in one point.
Then $|E\cap F|=\frac{96+126-112}{2}=55.$
By Lemma~\ref{lem:tool3-two-triangle-values}, both \(|C\cap E|\) and
\(|D\cap E|\) lie in \(S_6\).  If \(\varepsilon\in\{0,1\}\) records whether
the common point \(C\cap D\) lies in \(E\), then $|E\cap F|=|C\cap E|+|D\cap E|-2\varepsilon.$
Thus two elements of \(S_6\) would have to sum to \(55\) or \(57\), which is
impossible.  Hence the exceptional triple cannot occur.
\end{proof}

\subsection{Exclusion of the weight-\(126\) minimum-weight branch}
\label{app:a64-row-slice-summary}

We use the notation of Subsection~\ref{subsec:four-block-residuals}.  Thus
\(A,B\subseteq V\) are affine \(6\)-flats with \(|A\cap B|=1\),
\(s_{126}=\mathbf 1_A+\mathbf 1_B\), and \(S=\supp(s_{126})\).  For
\(g_3\in\RM(5,12)\), put
$\mu=|\supp(g_3)\cap S^c|, \rho=|\supp(g_3)\cap S|.$
If \(\tau=\wt(s_{126}+g_3)\in\{194,198,202,206\}\), then
$\tau=126+\mu-\rho,
\rho=\mu+126-\tau,
\wt(g_3)=2\mu+126-\tau .$

\begin{proposition}[Row-slice reduction for the weight-\(126\) branch]
\label{prop:a64-row-slice-reduction}
A target word in the family \(\mathcal F_{\rm A64}^{(126)}\) can occur only when
$\mu\in\{132,136,140\}.$
\end{proposition}

\begin{proof}
Let \(p=A\cap B\), and set
$K_A=|\supp(g_3)\cap A|,
K_B=|\supp(g_3)\cap B|,
\varepsilon_0=g_3(p).$
Since \(p\notin S\) but is counted once in each of \(K_A\) and \(K_B\),
$\rho=K_A+K_B-2\varepsilon_0.$
We compute the split weight support of \(\RM(5,12)\) with respect to the
partition
$\{p\}, A\setminus\{p\}, B\setminus\{p\},V\setminus(A\cup B).$
The computation retains the exact tuple
$(\varepsilon_0,K_A,K_B,\mu),$
so profiles with distinct values of these four statistics are not merged.
Filtering the resulting profiles by $K_A+K_B-2\varepsilon_0=\mu+126-\tau,
\tau\in\{194,198,202,206\},$
leaves only the three values \(\mu=132,136,140\).  This is an exact
enumeration of the split-profile set
$\{(\varepsilon_0,K_A,K_B,\mu) \mid g_3\in\RM(5,12)\}$, not a randomized search
or a search over selected representatives.
\end{proof}

\begin{lemma}[Low-weight input for \(\RM(5,12)\)]
\label{lem:a64-rm512-low-input}
The only nonzero weights of \(\RM(5,12)\) below \(224\) are \(128\) and
\(192\).  Moreover, every weight-\(192\) word in \(\RM(5,12)\) has the form
$\mathbf 1_C+\mathbf 1_D,$
where \(C,D\subseteq V\) are affine \(7\)-flats and \(C\cap D\) is an affine
\(5\)-flat.
\end{lemma}

\begin{proof}
By the Kasami--Tokura classification for \(\RM(5,12)\) below \(2\cdot128\),
the only nonzero weights below \(224\) are \(128\) and \(192\).  The
weight-\(192\) words are the \(\ell=2\) normal-form words, i.e. sums of two
minimum-weight words.  Thus they have the displayed form, and
\(192=128+128-2\cdot32\) forces \(C\cap D\) to be an affine \(5\)-flat.
\end{proof}

\subsubsection{The weight-\(192\) overlap verification}
\label{app:a64-weight192-verification}

\begin{proposition}[Weight-\(192\) overlap exclusion]
\label{prop:a64-weight192-overlap}
Let \(C,D\subseteq V\) be affine \(7\)-flats with \(C\cap D\) an affine
\(5\)-flat.  Then $|(C\triangle D)\cap S|\notin\{56,60\}.$
\end{proposition}

\begin{proof}
After translating the point \(A\cap B\) to the origin, write
\(V=A_0+ B_0\), where \(\dim A_0=\dim B_0=6\), and
$S=(A_0\setminus\{0\})\cup(B_0\setminus\{0\}).$
Let \(E=C\cap D\), and let \(E_0\) be its direction space.  Put
$a=\dim(E_0\cap A_0),
b=\dim(E_0\cap B_0),
k=5-a-b.$
The parameters \((a,b,k)\) determine the orbit of \(E_0\) under
\(GL(A_0)\times GL(B_0)\): indeed, \(E_0\) decomposes as
$(E_0\cap A_0)+(E_0\cap B_0)+\Gamma_\phi,$
where \(\Gamma_\phi\) is the graph of an isomorphism between two
\(k\)-dimensional subspaces of complementary quotients of \(A_0\) and
\(B_0\).

For each admissible direction type \((a,b,k)\), we enumerate the
affine cosets \(E=e+E_0\). For each such \(E\), the affine \(7\)-flats
\(C,D\) satisfying \(C\cap D=E\) correspond to pairs of
\(2\)-dimensional subspaces of \(V/E_0\) with trivial intersection,
after taking \(e+E_0\) as the origin in the quotient.
We enumerate all such quotient configurations and compute
$ |(C\triangle D)\cap S| $
exactly. For each admissible direction type, this procedure
yields the complete set of resulting overlap values.
Neither \(56\) nor \(60\) occurs in any of these finite sets.
This proves the proposition.
\end{proof}

\section{Punctured affine-flat geometry in the fixed shell}
\label{app:carlet-punctured-flat}

Throughout this appendix, use the notation of
Subsection~\ref{subsec:carlet-punctured}.  Thus
\(V=V_6\times V_6\), $Z=(V_6\setminus\{0\})\times(V_6\setminus\{0\}),$
and $H_1=V_6\times\{0\},
H_2=\{0\}\times V_6,
H=H_1\cup H_2.$
Hence \(Z=V\setminus H\).

\subsection{Punctured values of affine flats}
\label{app:carlet-punctured-value-sets}

\begin{lemma}[Possible sizes of \(X\cap H\)]
\label{lem:carlet-bad-shapes}
Let \(X\subseteq V\) be an affine \(r\)-flat.  Put
\(S_1=X\cap H_1\) and \(S_2=X\cap H_2\).  Then \(X\cap H=S_1\cup S_2\), and
\[
|X\cap H|\in
\{0\}\cup
\{2^s \mid 0\le s\le r\}
\cup
\{2^s+2^t-1 \mid s,t\ge0,\ s+t\le r\}
\cup
\{2^s+2^t \mid s,t\ge0,\ s+t+1\le r\}.
\]
\end{lemma}

\begin{proof}
Each nonempty \(S_i\) is an affine flat.  If only one side is nonempty, its
size is \(2^s\).  If both sides are nonempty, then they meet only possibly at
\(0\).  In the meeting case, the two direction spaces have trivial
intersection, giving \(s+t\le r\) and size \(2^s+2^t-1\).  In the disjoint
case, one additional direction joining the two flats is needed, giving
\(s+t+1\le r\) and size \(2^s+2^t\).
\end{proof}

\begin{proof}[Proof of Lemma~\ref{lem:punctured-values}]
For an affine \(r\)-flat \(X\), where \(r=5\) or \(r=3\), one has
$|X\cap Z|=2^r-|X\cap H|.$
Applying Lemma~\ref{lem:carlet-bad-shapes} gives
$|X\cap Z|\in \{0,15,16,21,22,23,24,25,26,27,28,29,30,31,32\}$
for \(r=5\), and $|X\cap Z|\in\{0,3,4,5,6,7,8\}$
for \(r=3\).
\end{proof}

\subsection{Two affine \(5\)-flat corrections}
\label{app:carlet-two-flat-cases}

\begin{lemma}[Coarse intersection-dimension exclusions]
\label{lem:carlet-coarse-intersection-cases}
Let \(A,B\subseteq V\) be distinct affine \(5\)-flats with
\(C=A\cap B\ne\varnothing\), and let $g=\mathbf 1_A+\mathbf 1_B.$
If $ \dim C\in\{0,1,2,4\},$
then $\wt(g|_Z)\notin T_{\rm red}.$
\end{lemma}

\begin{proof}
Put \(d=\dim C\).  In the full space, $\wt(g)=|A\triangle B|=64-2^{d+1}.$
If either \(|A\cap Z|\) or \(|B\cap Z|\) is zero, then \(g|_Z\) is the
restriction of a single affine \(5\)-flat indicator.  Its weight therefore
lies in \(P_5\), and hence not in \(T_{\rm red}\).

Otherwise, $|A\cap H|\le17$ and $|B\cap H|\le17,$
because the smallest positive element of \(P_5\) is \(15\). Thus $\wt(g|_H)\le34.$
For \(d=0,1,2\), this yields
\[
\begin{array}{c|c|c}
d & \wt(g) & \wt(g|_Z)\ge \wt(g)-34\\
\hline
0 & 62 & 28\\
1 & 60 & 26\\
2 & 56 & 22.
\end{array}
\]
These values cannot lie in \(T_{\rm red}\).

For \(d=4\), let $W=\operatorname{aff}(A\cup B)$
be the affine span of \(A\cup B\).  Since
\(\dim A=\dim B=5\), \(\dim(A\cap B)=4\), and \(A\ne B\), the space
\(W\) is an affine \(6\)-flat, and \(A\) and \(B\) are distinct affine
hyperplanes of \(W\). Over \(\mathbb F_2\), the
sum of their indicators is the indicator of another affine hyperplane
\(D\subseteq W\).  Hence \(\wt(g|_Z)=|D\cap Z|\in P_5\), again excluding
\(T_{\rm red}\).
\end{proof}

\begin{lemma}[Candidate screening for the \(d=3\) case]
\label{lem:carlet-d3-candidate-screening}
Assume \(g=\mathbf 1_A+\mathbf 1_B\), \(C=A\cap B\), and \(\dim C=3\).  Set
$a=|A\cap Z|, b=|B\cap Z|, c=|C\cap Z|.$
If \(a+b-2c\in T_{\rm red}\), then, up to interchanging \(A\) and \(B\), the
triple \((a,b,c)\) is one of those in Table~\ref{tab:carlet-d3-candidates}.
\end{lemma}

\begin{proof}
By Lemma~\ref{lem:punctured-values}, \(a,b\in P_5\) and \(c\in P_3\).  If
\(a=0\) or \(b=0\), then \(c=0\) and the reduced weight is a single element of
\(P_5\), already excluded.  Hence we may assume \(a,b>0\), and by symmetry
\(a\le b\).  Since \(c\le8\) and \(a+b-2c\le20\), we have
\(a+b\le36\).  Hence $(a,b)\in\{(15,15),(15,16),(16,16),(15,21)\}.$
Checking these four pairs against \(c\in P_3\) gives exactly
Table~\ref{tab:carlet-d3-candidates}.
\end{proof}

\begin{table}[htbp]
\centering
\caption{Candidate triples in the \(d=3\) case, where
\(a=|A\cap Z|\), \(b=|B\cap Z|\), and \(c=|A\cap B\cap Z|\).}

\small
\renewcommand{\arraystretch}{1.12}
\begin{tabular}{c p{0.44\textwidth} p{0.30\textwidth}}
\hline
Reduced target & Remaining triples \((a,b,c)\) &
Excluding local types \\
\hline
\(17\) & \((15,16,7)\) & type \(16\) \\
\(18\) & \((15,15,6),\ (16,16,7)\) & type \(15/16\) \\
\(19\) & \((15,16,6)\) & type \(15/16\) \\
\(20\) & \((15,15,5),\ (16,16,6),\ (15,21,8)\) & type \(15/16/21\) \\
\hline
\end{tabular}

\label{tab:carlet-d3-candidates}
\end{table}

\begin{lemma}[Local geometry of types \(15\), \(16\), and \(21\)]
\label{lem:carlet-type-geometry}
Let \(X\subseteq V\) be an affine \(5\)-flat.
\begin{enumerate}

\item If \(|X\cap Z|=15\), then, after possibly exchanging coordinate
blocks, $X\cap Z=S'\setminus\{p\},$
where \(S'\) is an affine \(4\)-flat of one of the forms
$S'=F\times\{v_0\}$ or $S'=\{u_0\}\times F,$
with \(F\subseteq V_6\) an affine \(4\)-flat and with \(v_0\ne0\) or
\(u_0\ne0\), respectively. The point \(p\) is the unique point of
\(S'\cap H\).
Hence every affine \(3\)-flat \(C\subseteq X\) satisfies
$|C\cap Z|\in\{0,3,4,7,8\}.$
Moreover, if \(C\subseteq X\cap Z\) is an affine \(3\)-flat, then one
coordinate block is constant on \(C\).

\item If \(|X\cap Z|=16\), then, after possibly exchanging coordinate
blocks, \(X\cap Z\) is an affine \(4\)-flat of the form
$F\times\{v_0\}$ or $\{u_0\}\times F,$
where \(F\subseteq V_6\) is an affine \(4\)-flat and the fixed coordinate
is nonzero.  Hence every affine \(3\)-flat \(C\subseteq X\) satisfies
$|C\cap Z|\in\{0,4,8\}.$

\item If \(|X\cap Z|=21\), then, after possibly exchanging coordinate blocks,
\(X=U\times W\), where \(U,W\le V_6\), \(\dim U=3\), \(\dim W=2\), and
$ X\cap Z=(U\setminus\{0\})\times(W\setminus\{0\}).$

Moreover, \(X\cap Z\) contains no affine \(3\)-flat on which one coordinate
block is constant.
\end{enumerate}
\end{lemma}

\begin{proof}
For type \(15\), \(|X\cap H|=17\).  In
Lemma~\ref{lem:carlet-bad-shapes}, the value \(17\) can only arise as
\(16+1\) in the disjoint case or as \(16+2-1\) in the meeting case.  In both
cases, one side contributes an affine \(4\)-flat contained in \(H_1\) or
\(H_2\), and the other side contributes exactly one additional bad point.
Thus \(X\) is the union of two parallel affine \(4\)-flats, one contained in
\(H\) and the other contained in a block slice; hence
\(X\cap Z=S'\setminus\{p\}\). Intersecting an affine \(3\)-flat
with the \(4\)-flat \(S'\) gives sizes \(0,4,8\), and removing the possible
point \(p\) gives \(\{0,3,4,7,8\}\).  If \(C\subseteq X\cap Z\), then
\(C\subseteq S'\), so one coordinate block is constant.

For type \(16\), \(|X\cap H|=16\).  The shape alternatives in
Lemma~\ref{lem:carlet-bad-shapes} force \(X\cap H\) to be an affine \(4\)-flat
contained in one of \(H_1,H_2\). The complementary parallel \(4\)-flat is
\(X\cap Z\), and the same dimension count gives \(\{0,4,8\}\).

For type \(21\), \(|X\cap H|=11\), and the only possible shape is the meeting
case \(11=8+4-1\).  Thus \(X\) passes through \(0\), and
\(X\cap H_1=U\times\{0\}\) and \(X\cap H_2=\{0\}\times W\), with
\(\dim U=3\) and \(\dim W=2\), after possibly exchanging blocks.  Since
\(\dim X=5\), these two subspaces span \(X=U\times W\).  It remains to prove 
the final assertion. Suppose that
\(C\subseteq X\cap Z\) is an affine \(3\)-flat on which one coordinate block
is constant.

If the \(U\)-coordinate is constant on \(C\), then the direction space of
\(C\) is contained in \(\{0\}\times W\), which has dimension \(2\).  This
contradicts \(\dim C=3\).

If the \(W\)-coordinate is constant on \(C\), then the projection of \(C\)
onto \(U\) is an affine \(3\)-flat in the \(3\)-dimensional space \(U\).
It must therefore equal \(U\), and hence contains \(0\).  Consequently,
\(C\) contains a point of the form \((0,w_0)\), which does not belong to
\(Z\).  This again contradicts \(C\subseteq X\cap Z\).

Thus \(X\cap Z\) contains no affine \(3\)-flat on which either coordinate
block is constant.
\end{proof}

\begin{lemma}[Exclusion of the \(d=3\) candidates]
\label{lem:carlet-d3-local-exclusion}
Under the hypotheses of Lemma~\ref{lem:carlet-d3-candidate-screening}, none of
the triples in Table~\ref{tab:carlet-d3-candidates} can occur.
\end{lemma}

\begin{proof}
If one of \(A,B\) has punctured value \(16\), then
Lemma~\ref{lem:carlet-type-geometry} gives \(c\in\{0,4,8\}\), excluding
\(c=6,7\).  If one side has punctured value \(15\), the same lemma gives
\(c\in\{0,3,4,7,8\}\), excluding \(c=5,6\).  This eliminates all candidates
except \((15,21,8)\).

For \((15,21,8)\), the equality \(c=8\) implies
\(C\subseteq A\cap Z\) and \(C\subseteq B\cap Z\).  On the type-\(15\) side,
Lemma~\ref{lem:carlet-type-geometry} forces one coordinate block to be constant
on \(C\).  On the type-\(21\) side, no such affine \(3\)-flat is contained in
the punctured part.  This contradiction excludes the final candidate.
\end{proof}

\section{Row-profile and mixed-core analysis in the fixed shell}
\label{app:inner-mm-refinements}

This appendix supports Subsection~\ref{subsec:carlet-inner-mm}.  We keep the
notation $ U=V_6\setminus\{0\}, Z=U\times U $ and $ 
T_{\rm red}=\{17,18,19,20\}.$
All weights in this appendix are reduced weights on \(Z\).  The target
exclusions below follow from explicit algebraic and geometric arguments.
A finite profile calculation is retained only for the quantitative pure-cross
refinement in Proposition~\ref{prop:app-pure-cross}; that refinement is not
needed for the fixed-shell exclusion.

\subsection{Pure-cross factors and separated presentations}
\label{app:inner-mm-pure-separated}

For nonempty subsets \(I,J\subseteq[6]\), put
$ \mu_I(u)=\prod_{i\in I}u_i$ and $
\nu_J(v)=\prod_{j\in J}v_j.$
An \emph{allowed degree-seven cross factor} is
\begin{equation}
\label{eq:allowed-cross-factor}
 c_{I,J}(u,v)=\mu_I(u)\nu_J(v),
 \qquad
 I,J\ne\varnothing,
 \qquad |I|+|J|=7.
\end{equation}
For \(1\le m\le5\), let \(\mathscr C_m\) be the set of nonzero restrictions
to \(Z\) of sums $  P(u,v)=\sum_{s=1}^{m}c_{I_s,J_s}(u,v),$
where the ordered pairs \((I_s,J_s)\) are distinct.  Repeated factors may be
removed in pairs because all sums are over \(\mathbb F_2\).  Set
$ \mathscr C_{\le5}=\bigcup_{m=1}^{5}\mathscr C_m. $
We also write \(\mathscr C_{\le5}^{\rm vd}\) for the subfamily in which the
sets \(J_1,\ldots,J_m\) are pairwise distinct.  This last condition is the
normal form used in the quantitative profile calculation below; the target
exclusion itself applies to the full family \(\mathscr C_{\le5}\).

For \(0\le d\le6\), define the punctured Reed--Muller space
$ \mathcal R_d=\{f|_U \mid f\in\RM(d,6)\}, \quad \mathcal R_{-1}=\{0\}.$
The restriction map is injective for \(d\le5\): a nonzero Boolean
function supported only at the origin has algebraic degree six.  Since
\(\dim\RM(5,6)=63=|U|\), it follows that
$ \mathcal R_5=\mathcal R_6=\mathbb F_2^U. $
For \(q:U\to\mathbb F_2\), recall that
$ \delta(q)=\min\{\wt_U(q),63-\wt_U(q)\}. $

\begin{lemma}[Punctured local spectra]
\label{lem:punctured-local-spectra}
The following facts hold.
\begin{enumerate}
\item For \(1\le d\le5\), every nonzero word in \(\mathcal R_d\) has
weight at least $ \Delta_d=2^{6-d}-1,$
and this value is attained.
\item If \(q\in\mathcal R_d\setminus\mathcal R_{d-1}\), then
$  \delta(q)\ge \mu_d:=2^{6-d}-1, \quad 1\le d\le5.$
Thus $(\mu_1,\mu_2,\mu_3,\mu_4,\mu_5)=(31,15,7,3,1).$
\item The exact positive \(\delta\)-spectrum of \(\mathcal R_2\) is
\begin{equation}
\label{eq:punctured-rm2-delta-spectrum}
\{\delta(q) \mid q\in\mathcal R_2,\ \delta(q)>0\}
=
\{15,16,23,24,27,28,31\}.
\end{equation}
In particular, the only nonzero weights of words in \(\mathcal R_2\) not
exceeding \(20\) are \(15\) and \(16\).
\end{enumerate}
\end{lemma}

\begin{proof}
The minimum distance of \(\RM(d,6)\) is \(2^{6-d}\).  Puncturing at the zero
vector decreases a nonzero weight by at most one, and equality is attained by
\(\prod_{i=1}^{d}(1+x_i)\), whose support is a linear subspace containing
zero.  This proves the first assertion.  Since both \(q\) and \(q+1\) have
algebraic degree exactly \(d\), the same bound applied to the lighter of the
two proves the second assertion.

For the third assertion, use the standard rank classification of quadratic
Boolean functions in six variables.  If the quadratic part has rank
\(2r\in\{2,4,6\}\), then the full-space weight is either \(32\) or
\(32\pm2^{5-r}\); the balanced value \(32\) occurs when the affine part is
nontrivial on the radical.  Puncturing at zero changes the weight by zero or
one.  The unbalanced cases therefore give 
$ \{15,16\}, \{23,24\} $ and $ \{27,28\}$ respectively
for the positive values of \(\delta\), while every balanced nonconstant
quadratic or affine function gives \(\delta=31\).  Constants give
\(\delta=0\).  This proves \eqref{eq:punctured-rm2-delta-spectrum} and the
last statement.
\end{proof}

A \emph{degree-compatible separated presentation} on \(Z\) is an expression
\begin{equation}
\label{eq:separated-presentation}
P(u,v)=\sum_{i=1}^{r}\alpha_i(u)q_i(v),
\qquad
\alpha_i\in\mathcal R_{a_i},\quad
q_i\in\mathcal R_{b_i},\quad
1\le b_i\le5,\quad a_i+b_i\le7.
\end{equation}
Terms depending only on \(u\) are absorbed into the row function \(h(u)\).
The restriction to \(U\) of a degree-six \(v\)-monomial can also be replaced
by a polynomial of degree at most five: on \(U\),
$ \prod_{j=1}^{6}(1+v_j)=0. $
This replacement does not increase the total degree.  Thus every restriction
to \(Z\) of a function in \(\RM(7,12)\) has a presentation of the form
\eqref{eq:separated-presentation}.

To state the two- and three-block ranges without choosing canonical
representatives, define the reduced block rank by
\begin{equation}
\label{eq:reduced-block-rank}
\operatorname{br}_7(P)
=
\min\left\{
 r \mid 
 P=a(u)+\sum_{i=1}^{r}\alpha_i(u)q_i(v)
 \text{ on }Z,
 \text{ with the constraints in \eqref{eq:separated-presentation}}
\right\}.
\end{equation}
We use the intrinsic families
\begin{equation}
\label{eq:L2-L3-definition}
\mathscr L_2=\{P\mid\operatorname{br}_7(P)\le2\},
\qquad
\mathscr L_3=\{P\mid\operatorname{br}_7(P)\le3\}.
\end{equation}
They are conservative enlargements of any choice of canonical two- or
three-block representatives, so an exclusion for \(\mathscr L_2\) or
\(\mathscr L_3\) also applies to every such reduced representative family.

More precisely, let \(D\) be the least integer such that every row
\(P(u,\cdot)\) lies in \(\mathcal R_D\).  Modulo
\(\mathcal R_{D-1}\), the row map defines an intrinsic tensor
\begin{equation}
\label{eq:top-layer-tensor}
\Theta_D(P)
\in
\mathcal R_{7-D}\otimes
\bigl(\mathcal R_D/\mathcal R_{D-1}\bigr).
\end{equation}
Its tensor rank, denoted by \(\rho_D(P)\), is the highest-layer block rank.
This is the precise rank reduction used below and is independent of all
choices of bases or representatives.

\begin{proposition}[Low-weight row gap]
\label{prop:app-separated}
Let $ g(u,v)=h(u)+P(u,v), $
where \(h:U\to\mathbb F_2\) is arbitrary and \(P\) has a
degree-compatible separated presentation of total degree at most \(7\).
Then $ \wt(g)\notin T_{\rm red}.$
In particular, this conclusion applies to \(\mathscr L_2\), \(\mathscr L_3\),
the pure-cross family \(\mathscr C_{\le5}\), the base families defined below,
and sums of at most three affine \(5\)-flat indicators.
\end{proposition}

\begin{proof}
Assume, for contradiction, that \(\wt(g)\le20\).  For a fixed \(u\), put
\(w_u=\wt_U(P(u,\cdot))\).  The contribution of the row \(u\) to \(g\) is
either \(w_u\) or \(63-w_u\).  The larger of these two numbers is at least
\(32\).  Hence every row of a word of total weight at most \(20\) must use its
smaller contribution, and therefore
\begin{equation}
\label{eq:low-weight-row-rigidity}
\wt(g)=\sum_{u\in U}\delta(P(u,\cdot)).
\end{equation}
This observation upgrades the row-wise minimization in
Lemma~\ref{lem:row-wise-elimination} to an identity for every putative target
word.

If \(P\) has no nonconstant \(v\)-row, the right-hand side of
\eqref{eq:low-weight-row-rigidity} is zero.  We may therefore assume that its
largest visible \(v\)-degree \(D\) satisfies \(1\le D\le5\).

Suppose first that \(D\le2\).  Every nonconstant row belongs to
\(\mathcal R_2\), so by
Lemma~\ref{lem:punctured-local-spectra} its positive local contribution lies
in $ \{15,16,23,24,27,28,31\}.$
If there are two nonconstant rows, the total is at least \(30\); if there is
only one, the total is one of the displayed values.  Neither possibility lies
in \(T_{\rm red}\).

Now let \(D=3\) or \(4\).  Choose a nonzero coefficient function \(\lambda\)
of the tensor \(\Theta_D(P)\).  Then
\(\lambda\in\mathcal R_{7-D}\).  Every row in \(\supp(\lambda)\) has a
nonzero degree-\(D\) quotient and hence contributes at least \(\mu_D\).
Lemma~\ref{lem:punctured-local-spectra} gives
\[
\wt(g)\ge
\begin{cases}
3\cdot7=21,&D=3,\\
7\cdot3=21,&D=4.
\end{cases}
\]

It remains to consider \(D=5\).  Put \(k=\rho_5(P)\).  Choose bases in
\eqref{eq:top-layer-tensor} so that the top layer is represented by
$ \sum_{j=1}^{k}\lambda_j(u)\,\overline q_j(v), $
where \(\lambda_1,\ldots,\lambda_k\in\mathcal R_2\) are linearly independent
and \(\overline q_1,\ldots,\overline q_k\) are linearly independent in
\(\mathcal R_5/\mathcal R_4\).

If \(k\ge2\), take \(\lambda_1,\lambda_2\).  The three nonzero words
\(\lambda_1,\lambda_2,\lambda_1+\lambda_2\) have weights at least \(15\).
Consequently,
$ 2\,|\supp(\lambda_1)\cup\supp(\lambda_2)|=\wt(\lambda_1)+\wt(\lambda_2)+\wt(\lambda_1+\lambda_2)
\ge45.$
Thus at least \(23\) rows have a nonzero degree-five quotient, and each such
row contributes at least \(\mu_5=1\).  Hence \(\wt(g)\ge23\).

Finally, suppose \(k=1\).  The degree-five layer has the form
\(\lambda(u)\overline q(v)\), with
\(0\ne\lambda\in\mathcal R_2\).  The code \(\mathcal R_4\) is the binary
\([63,57,3]\) Hamming code: it has dimension \(57\), minimum distance \(3\),
and meets the Hamming bound with equality.  Hence every nonzero coset of
\(\mathcal R_4\) in \(\mathcal R_5=\mathbb F_2^U\) has a unique unit-vector
representative.  After changing the representative of \(\overline q\) and
absorbing the resulting degree-at-most-four term into the lower layer, we may
write
\begin{equation}
\label{eq:rank-one-degree-five-normal-form}
P(u,v)=\lambda(u)\rho(v)+R(u,v),
\end{equation}
where \(\rho\) is supported at one point of \(U\) and every row of \(R\)
belongs to \(\mathcal R_4\).  Any \(v\)-constant part of \(R\) is absorbed
into \(h\).  Put \(B=\supp(\lambda)\), so \(|B|\ge15\).

If \(R=0\), then the right-hand side of
\eqref{eq:low-weight-row-rigidity} equals \(|B|\).  By the quadratic
classification in Lemma~\ref{lem:punctured-local-spectra}, a nonzero word of
\(\mathcal R_2\) of weight at most \(20\) has weight \(15\) or \(16\), not a
member of \(T_{\rm red}\).

Assume that \(R\ne0\), and let \(d\in\{1,2,3,4\}\) be its largest visible
\(v\)-degree.  Choose a nonzero coefficient function
\(\sigma\in\mathcal R_{7-d}\) in that layer, and put
\(S=\supp(\sigma)\).  Then
\[
|S|\ge
\begin{cases}
1,&d=1,2,\\
3,&d=3,\\
7,&d=4.
\end{cases}
\]
For \(u\in S\setminus B\), the row cost is at least \(\mu_d\).  For
\(u\in S\cap B\), changing a word at the one coordinate supporting \(\rho\)
changes its distance to the two constant words by at most one, and hence
$ \delta(\rho+R(u,\cdot))\ge\mu_d-1.$
For every \(u\in B\), the row cost is at least one, because the nonzero coset
\(\rho+\mathcal R_4\) contains neither constant word.  Relative to this
baseline contribution \(|B|\), we obtain
\[
\wt(g)
\ge
|B|+(\mu_d-2)|S\cap B|+\mu_d|S\setminus B|
\ge
|B|+(\mu_d-2)|S|.
\]
For \(d=1,2,3,4\), the last additional term is at least, respectively,
$ 29, 13, 15, 7.$
Therefore \(\wt(g)\ge15+7=22\), completing the proof.
\end{proof}

We next specify the two normalized low-weight base configurations used in
the perturbation analysis.  Put
\begin{equation}
\label{eq:explicit-15-base}
\alpha_*(u)=\prod_{i=1}^{5}(1+u_i),
\qquad
q_0(v)=v_1+v_2+v_1v_2,
\qquad
B_{15}(u,v)=\alpha_*(u)(1+q_0(v)).
\end{equation}
On \(U\), \(\alpha_*\) is supported at \(e_6\), while
\(1+q_0=(1+v_1)(1+v_2)\) has weight \(15\); hence \(B_{15}\) has reduced
weight \(15\).  Also put
\begin{equation}
\label{eq:explicit-16-base}
\alpha_{16}(u)=u_1u_2,
\qquad
\rho_*(v)=\prod_{j=1}^{5}(1+v_j),
\qquad
B_{16}(u,v)=\alpha_{16}(u)\rho_*(v).
\end{equation}
The restriction of \(\rho_*\) to \(U\) is a one-point function, so
\(B_{16}\) has reduced weight \(16\).

For clarity, define the corresponding perturbation families by
\begin{align}
\mathscr F_{15}^{(\le3)}
&=
\left\{
 h+B_{15}+\sum_{i=1}^{r}\beta_i(u)r_i(v)\mid
 \begin{array}{l}
 0\le r\le3,\quad
 \beta_i\in\mathcal R_{a_i},\quad r_i\in\mathcal R_{b_i},\\[-1mm]
 1\le b_i\le5,\quad a_i+b_i\le7
 \end{array}
\right\},
\label{eq:F15-family}\\
\mathscr F_{16}^{(<5)}
&=
\left\{
 h+B_{16}+P \mid
 \begin{array}{l}
 P\text{ has a presentation \eqref{eq:separated-presentation}},\\[-1mm]
 \text{and every payload has }v\text{-degree at most }4
 \end{array}
\right\}.
\label{eq:F16-family}
\end{align}
Here \(h:U\to\mathbb F_2\) is arbitrary.  These explicitly defined families are conservative enlargements of the
weight-\(15\) and weight-\(16\) normal-form families considered here.
Hence an exclusion for the larger families also excludes those normal forms.  Proposition~\ref{prop:app-separated} immediately
excludes \(T_{\rm red}\) from both families.

\begin{proposition}[Pure-cross exclusion and quantitative refinement]
\label{prop:app-pure-cross}
For every \(P\in\mathscr C_{\le5}\) and every row function \(h\),
$ \wt(h+P)\notin T_{\rm red}.$
Moreover, for the \(v\)-distinct profile subfamily
\(\mathscr C_{\le5}^{\rm vd}\),
$ \min_{h:U\to\mathbb F_2}\wt(h+P)\ge31, $
and equality occurs.
\end{proposition}

\begin{proof}
The target exclusion is a direct application of
Proposition~\ref{prop:app-separated}.  We record the quantitative
\(v\)-distinct refinement because it explains the observed \(31\)-barrier in
the pure-cross search.

For an ordered tuple \({\bf I}=(I_1,\ldots,I_m)\) and
\(A\subseteq[m]\), define
$ n_A({\bf I})=\#\left\{u\in U\mid
\{s\mid I_s\subseteq\supp(u)\}=A
\right\}.$
For an ordered tuple \({\bf J}=(J_1,\ldots,J_m)\), define
$ \lambda_A({\bf J})=\delta\!\left(\sum_{s\in A}\nu_{J_s}\right).$
Lemma~\ref{lem:row-wise-elimination} gives the exact finite profile formula
\begin{equation}
\label{eq:pure-cross-profile-formula}
\min_h\wt(h+P)
=
\sum_{A\subseteq[m]}n_A({\bf I})\lambda_A({\bf J}).
\end{equation}
For \(m\le4\), direct evaluation of
\eqref{eq:pure-cross-profile-formula} over the complete coordinate-permutation
profiles gives minimum \(31\).

For \(m=5\), encode an ordered \({\bf I}\)-profile by
$ c_\tau
=
\#\left\{i\in[6] \mid
(1_{i\in I_1},\ldots,1_{i\in I_5})=\tau
\right\}, \tau\in\{0,1\}^5.$
There are \(2{,}069{,}818\) such ordered profiles with all five \(I_s\)
nonempty, of which \(1{,}872{,}011\) admit a degree-compatible tuple of five
pairwise distinct \(J_s\).  Put \(d_s=7-|I_s|\).  For a degree multiset
\(D=(d_1,\ldots,d_t)\), \(1\le t\le5\), set
\[
\varrho(D)
=
\min\left\{
\delta\!\left(\sum_{i=1}^{t}\nu_{K_i}\right) \mid
|K_i|=d_i,\ K_1,\ldots,K_t\text{ pairwise distinct}
\right\},
\qquad
\varrho(\varnothing)=0.
\]
Then $ L({\bf I})=\sum_{A\subseteq[5]}
 n_A({\bf I})\,\varrho\bigl((d_s)_{s\in A}\bigr)$
is a lower bound for every compatible \({\bf J}\)-profile.  The complete
integer-profile sieve leaves exactly eleven ordered profiles with
\(L({\bf I})\le31\), and none with \(L({\bf I})<31\).  Up to permutations of
the five summands and of the \(u\)-coordinates, they reduce to
$ I_1=\cdots=I_5=[6]$
and $ I_1=I_2=I_3=[6],I_4=I_5=T,|T|=5.$
Their lower bounds are, respectively,
$ \varrho(1,1,1,1,1)=31$
and $ \varrho(1,1,1,2,2)+\varrho(2,2)=15+16=31.$
Thus the minimum is at least \(31\).  Equality is attained by
$ P(u,v)=\mu_{[6]}(u)(v_1+v_2+v_3+v_4+v_5),h(u)=\mu_{[6]}(u),$
for which only the row \(u=(1,\ldots,1)\) is nonzero and has punctured weight
\(31\).
\end{proof}

\subsection{Mixed affine cores}
\label{app:inner-mm-mixed-affine}

Let \(S=a+L\subseteq V_6\times V_6\) be an affine \(5\)-flat and put
\(X=S\cap Z\).  Let $ H_1=V_6\times\{0\} $ and $ H_2=\{0\}\times V_6,$
and define $r_1=\dim(L\cap H_1),r_2=\dim(L\cap H_2)$ and $ k=5-r_1-r_2.$
We call \(X\) genuinely mixed if \(k>0\).  Let \(\epsilon_i=1\) if
\(S\cap H_i\ne\varnothing\) and \(\epsilon_i=0\) otherwise, and let
\(\epsilon_0=1\) if \(0\in S\) and \(\epsilon_0=0\) otherwise.  Then
\begin{equation}
\label{eq:mixed-core-size}
|S\cap Z|
=
32-\epsilon_1 2^{r_1}-\epsilon_2 2^{r_2}+\epsilon_0.
\end{equation}

\begin{lemma}[Single-core values]
\label{lem:single-core-values}
For a genuinely mixed core, the possible values of \(|S\cap Z|\) not
exceeding \(24\) are contained in $\{15,16,22,23,24\}.$
In particular, no single genuinely mixed core has reduced weight in
\(T_{\rm red}\).
\end{lemma}

\begin{proof}
Since \(k=5-r_1-r_2>0\), one has \(r_1+r_2\le4\).  Moreover,
\(\epsilon_0=1\) implies \(\epsilon_1=\epsilon_2=1\).  Substitution in
\eqref{eq:mixed-core-size} gives only the displayed values up to \(24\).
\end{proof}

\subsection{Three-core corrections}
\label{app:inner-mm-three-core}

Define the full mathematical family
\begin{equation}
\label{eq:M3-definition}
\begin{aligned}
\mathscr M_3=\bigl\{(X_1,X_2,X_3)\mid\;&
X_i=S_i\cap Z,\quad S_i=a_i+L_i\subseteq V_6\times V_6,\\
&\dim L_i=5,\quad
5-\dim(L_i\cap H_1)-\dim(L_i\cap H_2)>0,\\
& i=1,2,3\bigr\}.
\end{aligned}
\end{equation}
Thus \(\mathscr M_3\) is the full family of all ordered triples of
genuinely mixed cores.  For a member of
\(\mathscr M_3\), put $ a_{ij}=|X_i\cap X_j| $ and $
t=|X_1\cap X_2\cap X_3|.$
Then
\begin{equation}
\label{eq:three-core-symmetric-difference}
|X_1\triangle X_2\triangle X_3|
=
\sum_{i=1}^{3}|X_i|-2(a_{12}+a_{13}+a_{23})+4t.
\end{equation}

\begin{lemma}[Arithmetic of the two lowest residual patterns]
\label{lem:three-core-low-end-reduction}
The following conditional implications hold.
\begin{enumerate}
\item If $ (|X_1|,|X_2|,|X_3|)=(22,23,23),
\quad a_{ij}\le8,$
and \(|X_1\triangle X_2\triangle X_3|=20\), then
$ a_{12}=a_{13}=a_{23}=8,
\quad t=0.$
\item If $ (|X_1|,|X_2|,|X_3|)=(22,22,24)$
and \(|X_1\triangle X_2\triangle X_3|=18\), then
$ a_{12}+a_{13}+a_{23}-2t=25.$
\end{enumerate}
\end{lemma}

\begin{proof}
Both statements are immediate rearrangements of
\eqref{eq:three-core-symmetric-difference}.  In the first case,
$ a_{12}+a_{13}+a_{23}-2t=24.$
The three upper bounds \(a_{ij}\le8\), together with \(t\ge0\), force equality
in all three bounds and then \(t=0\).  The second identity follows from total
single-core size \(68\) and target weight \(18\).
\end{proof}

\begin{proposition}[Three-core exclusion in the fixed shell]
\label{prop:three-core-residual-exclusions}
For every \((X_1,X_2,X_3)\in\mathscr M_3\),
\[
|X_1\triangle X_2\triangle X_3|
\notin T_{\rm red}.
\]
\end{proposition}

\begin{proof}
The indicator of an affine \(5\)-flat in the \(12\)-dimensional ambient space
is a product of seven independent affine linear equations and therefore
belongs to \(\RM(7,12)\).  Hence
$ g=\mathbf 1_{S_1}+\mathbf 1_{S_2}+\mathbf 1_{S_3}
\in\RM(7,12).$
Its restriction to \(Z\) has a degree-compatible separated presentation, so
Proposition~\ref{prop:app-separated} applies.  Since
$ \supp(g|_Z)=X_1\triangle X_2\triangle X_3,$
the claimed exclusion follows.  No orbit count or finite candidate list is
used in this argument.
\end{proof}

\begin{remark}[Independent check for the \((22,22,24)\) pattern]
The second pattern in Lemma~\ref{lem:three-core-low-end-reduction} also has a
direct affine-geometric exclusion.  If every full affine intersection
\(S_i\cap S_j\) has dimension at most \(3\), then every \(a_{ij}\le8\), so
$ a_{12}+a_{13}+a_{23}-2t\le24.$
If some pair has affine intersection dimension \(4\), the two distinct
affine \(5\)-flats are affine hyperplanes in a common affine \(6\)-flat, and
their symmetric difference is a third affine \(5\)-flat \(D\).  If
\(D=S_k\), the three indicators sum to zero.  If \(D\ne S_k\), the correction
is the sum of two distinct affine \(5\)-flat indicators and is excluded by
Proposition~\ref{prop:fixed-shell-affine-flat-exclusion}.  Coincident reduced
supports cancel on \(Z\).  Thus the value \(25\) required by
Lemma~\ref{lem:three-core-low-end-reduction} cannot occur.
\end{remark}

\section{Profile analysis of sparse sums of degree-\(7\) coordinate monomials}
\label{app:sparse7-monomial-sums}

Let \([14]=\{1,\ldots,14\}\), and put $ T=\{322,326,330,334\}.$
For \(I\in\binom{[14]}{7}\), write $ m_I=\prod_{\ell\in I}x_\ell.$
The support of \(m_I\) is $ A_I=\{x\in\mathbb F_2^{14}\mid x_\ell=1\text{ for all }\ell\in I\},$
and \(|A_I|=128\).  For \(s\ge0\), define
\[
        \mathcal W_s^{\rm dist}
        =
        \left\{
        \wt\left(\sum_{I\in\mathcal I}m_I\right)\mid
        \mathcal I\subseteq\binom{[14]}{7},\ |\mathcal I|=s
        \right\}.
\]
All sums are over \(\mathbb F_2\).

\subsection{The symmetric-difference weight formula}
\label{app:sparse7-symmetric-difference}

\begin{lemma}[Symmetric-difference formula]
\label{lem:sparse7-symmetric-difference}
Let \(s\ge1\), and let
\(I_1,\ldots,I_s\in\binom{[14]}{7}\) be distinct.  Then
\[
\wt\left(\sum_{r=1}^s m_{I_r}\right)
=
\sum_{\emptyset\ne J\subseteq[s]}
(-2)^{|J|-1}
2^{14-\left|\bigcup_{j\in J}I_j\right|}.
\]
\end{lemma}

\begin{proof}
The support of the sum is
\(\triangle_{r=1}^s A_{I_r}\).  For arbitrary sets \(A_1,\ldots,A_s\),
\[
        \left|\triangle_{i=1}^s A_i\right|
        =
        \sum_{\emptyset\ne J\subseteq[s]}
        (-2)^{|J|-1}
        \left|\bigcap_{j\in J}A_j\right|.
\]
For monomial supports,
\[
        \left|\bigcap_{j\in J}A_{I_j}\right|
        =
        2^{14-\left|\bigcup_{j\in J}I_j\right|}.
\]
Substitution gives the formula.
\end{proof}

\subsection{Exact low-term analysis for \(0\le s\le4\)}
\label{app:sparse7-low-term}

\begin{lemma}[Two coordinate degree-seven monomials]
\label{lem:sparse7-two-term}
Let \(I,J\in\binom{[14]}{7}\) be distinct and set \(c=|I\cap J|\).  Then
$ \wt(m_I+m_J)=256-2^{c+1}.$
Consequently, $ \mathcal W_2^{\rm dist}=\{128,192,224,240,248,252,254\}.$
\end{lemma}

\begin{proof}
The support is \(A_I\triangle A_J\).  Since
$ |A_I\cap A_J|=2^{14-|I\cup J|}=2^c,$
we obtain $ \wt(m_I+m_J)=128+128-2\cdot2^c=256-2^{c+1}. $
For distinct \(I,J\), one has \(0\le c\le6\).
\end{proof}

\begin{proposition}[Exact low-term exclusion]
\label{prop:sparse7-low-term}
For \(0\le s\le4\), $ \mathcal W_s^{\rm dist}\cap T=\varnothing.$
\end{proposition}

\begin{proof}
The cases \(s=0\) and \(s=1\) give weights \(0\) and \(128\).  The case
\(s=2\) follows from Lemma~\ref{lem:sparse7-two-term}.  For \(s=3,4\), use the
Venn-region variables
\[
        n_J
        =
        \left|
        \left(\bigcap_{j\in J}I_j\right)
        \setminus
        \left(\bigcup_{j\notin J}I_j\right)
        \right|,
        \qquad
        J\subseteq[s].
\]
They satisfy $ \sum_{J\subseteq[s]}n_J=14,
\sum_{J\ni r}n_J=7\quad (r=1,\ldots,s),$ where $ n_J\ge0.$
The condition that the monomials are distinct is imposed by
\[
        \sum_{\substack{J\subseteq[s]\\ |\{i,j\}\cap J|=1}} n_J>0
        \qquad
        (i\ne j),
\]
which is equivalent to \(I_i\ne I_j\).  Conversely, every nonnegative integer
profile satisfying these conditions is realizable by assigning the fourteen
coordinates to the corresponding Venn regions.  Moreover, for every nonempty
\(K\subseteq[s]\),
\[
        \left|\bigcup_{k\in K} I_k\right|
        =
        \sum_{\substack{J\subseteq[s]\\ J\cap K\ne\varnothing}} n_J.
\]
Thus Lemma~\ref{lem:sparse7-symmetric-difference} determines the weight
directly from the Venn tuple.  Enumerating all feasible tuples gives exact
weight sets for \(s=3\) and \(s=4\), neither of which intersects \(T\).
\end{proof}

\subsection{Profile enumeration for \(5\le s\le7\)}
\label{app:sparse7-enumeration}

\begin{proposition}[Profile exclusion for \(5\le s\le7\)]
\label{prop:sparse7-s5s7-enumeration}
For \(s=5,6,7\), $ \mathcal W_s^{\rm dist}\cap T=\varnothing.$
\end{proposition}

\begin{proof}
For an ordered \(s\)-tuple \((I_1,\ldots,I_s)\), define its Venn-region profile
by
\[
        n_J
        =
        \left|
        \left(\bigcap_{j\in J}I_j\right)
        \setminus
        \left(\bigcup_{j\notin J}I_j\right)
        \right|,
        \qquad
        J\subseteq[s].
\]
The profile satisfies $ \sum_{J\subseteq[s]}n_J=14,
\sum_{J\ni r}n_J=7\quad (r=1,\ldots,s),$ where $ n_J\ge0,$
together with the pairwise-distinctness conditions
\[
        \sum_{\substack{J\subseteq[s]\\ |\{i,j\}\cap J|=1}}n_J>0
        \qquad (i\ne j).
\]
Conversely, every nonnegative integer profile satisfying these conditions is
realizable by assigning coordinates to Venn regions. Moreover, two ordered \(s\)-tuples 
of subsets lie in the same orbit under the natural action of \(S_{14}\) on the coordinates 
if and only if they have the same Venn-region profile. For every nonempty
\(K\subseteq[s]\),
\[
        \left|\bigcup_{k\in K}I_k\right|
        =
        \sum_{\substack{J\subseteq[s]\\ J\cap K\ne\varnothing}}n_J.
\]
Hence the weight is determined entirely by the profile through
Lemma~\ref{lem:sparse7-symmetric-difference}.

We enumerate all feasible profiles modulo the natural action of \(S_s\)
permuting the summands.  Since the Venn-region profile completely determines
the \(S_{14}\)-orbit of an ordered tuple, this gives an exact enumeration of 
the \(S_{14}\)-orbits in the full unordered distinct-summand family for each \(s=5,6,7\). 
None of the resulting weights lies in \(T\).
\end{proof}

\subsection{Complementary-pair reduction for \(s=8\)}
\label{app:sparse7-fixed-pair}

\begin{lemma}[Complementary-pair reduction modulo \(4\)]
\label{lem:sparse7-fixed-pair}
Let $ f=\sum_{r=1}^{8}m_{I_r},$
where \(I_1,\ldots,I_8\) are distinct \(7\)-subsets of \([14]\).  If
$ \wt(f)\equiv2\pmod4,$
then some pair \(I_i,I_j\) is disjoint.  Consequently, after permuting
variables and reordering summands, one may assume
$ I_1=\{1,\ldots,7\},I_2=\{8,\ldots,14\}.$
\end{lemma}

\begin{proof}
Modulo \(4\), the formula in Lemma~\ref{lem:sparse7-symmetric-difference}
reduces to the pairwise part: $ \wt(f)
\equiv
-2\sum_{1\le i<j\le8}|A_{I_i}\cap A_{I_j}|
\pmod4.$
For \(7\)-subsets \(I_i,I_j\), $ |A_{I_i}\cap A_{I_j}|=
2^{14-|I_i\cup I_j|}=2^{|I_i\cap I_j|}.$
This number is odd exactly when \(I_i\cap I_j=\varnothing\). Hence
$ \wt(f)\equiv2D\pmod4,$
where \(D\) is the number of disjoint pairs among
\(I_1,\ldots,I_8\).  If \(\wt(f)\equiv2\pmod4\), then \(D\) is odd, so a
disjoint pair exists.  Two disjoint \(7\)-subsets of \([14]\) are complementary,
and \(S_{14}\) acts transitively on complementary pairs.  The normalization
follows.
\end{proof}

\begin{proposition}[Fixed-pair profile exclusion for \(s=8\)]
\label{prop:sparse7-fixed-pair-enumeration}
For \(s=8\), $ \mathcal W_8^{\rm dist}\cap T=\varnothing.$
\end{proposition}

\begin{proof}
All weights in \(T\) are congruent to \(2\pmod4\).  By
Lemma~\ref{lem:sparse7-fixed-pair}, every target configuration contains a
complementary pair.  After permuting variables and reordering the summands,
normalize this pair as $ I_1=\{1,\ldots,7\}, I_2=\{8,\ldots,14\}.$

We then enumerate feasible Venn-region profiles of
\((I_1,\ldots,I_8)\) subject to this fixed-pair condition. Thus
$ n_J=0 \text{ whenever } |J\cap\{1,2\}|\ne1,$
and the remaining \(n_J\)'s satisfy
$\sum_{J\subseteq[8]}n_J=14,
\sum_{J\ni r}n_J=7
\quad (r=1,\ldots,8), $ where $
n_J\ge0,$
together with the pairwise-distinctness conditions
\[
        \sum_{\substack{J\subseteq[8]\\ |\{i,j\}\cap J|=1}}n_J>0
        \qquad (i\ne j).
\]
For every nonempty \(K\subseteq[8]\),
\[
        \left|\bigcup_{k\in K}I_k\right|
        =
        \sum_{\substack{J\subseteq[8]\\ J\cap K\ne\varnothing}}n_J,
\]
so the weight is again determined by
Lemma~\ref{lem:sparse7-symmetric-difference}.

The Venn-region profile already identifies configurations that differ only
by permutations of the coordinates within the two fixed seven-element
blocks. In addition, we identify profiles under the natural action of
\(S_6\) permuting the remaining summands \(I_3,\ldots,I_8\). 
The resulting fixed-pair profile list contains no weight in \(T\).  Hence
\(\mathcal W_8^{\rm dist}\cap T=\varnothing\).
\end{proof}

\subsection{Repeated summands}
\label{app:sparse7-repetitions}

If repetitions are allowed in $ \sum_{r=1}^t m_{I_r}, $
then equal monomials cancel in pairs over \(\mathbb F_2\).  The expression
therefore reduces to a sum of \(s\le t\) distinct monomials.  Combining
Propositions~\ref{prop:sparse7-low-term},
\ref{prop:sparse7-s5s7-enumeration}, and
\ref{prop:sparse7-fixed-pair-enumeration}, we obtain
\[
        \wt\left(\sum_{r=1}^t m_{I_r}\right)\notin T
        \qquad
        (t\le8).
\]
This conclusion is confined to sums of squarefree coordinate monomials of
degree \(7\).

\end{document}